\documentclass[acmtog,nonacm]{acmart}

\renewcommand\footnotetextcopyrightpermission[1]{}
\usepackage{booktabs} 

\usepackage[ruled]{algorithm2e} 

\SetAlFnt{\small}
\SetAlCapFnt{\small}
\SetAlCapNameFnt{\small}
\SetAlCapHSkip{0pt}
\usepackage{algpseudocode}
\usepackage{xcolor}

\usepackage[table]{xcolor} 

\acmJournal{TOG}

\begin{document}

\newcommand{\name}{\textit{AniGen}\xspace}

\newcommand{\revise}[1]{\textcolor{black}{#1}}

\title{AniGen: Unified $S^3$ Fields for Animatable 3D Asset Generation}

\author{Yi-Hua Huang}
\affiliation{\institution{The University of Hong Kong}\city{China}\country{China}}
\authornote{This work is in collaboration with VAST.}

\author{Zi-Xin Zou}
\affiliation{\institution{VAST}\country{China}}

\author{Yuting He}
\affiliation{\institution{The Chinese University of Hong Kong}\country{China}}

\author{Chirui Chang}
\affiliation{\institution{The University of Hong Kong}\country{China}}

\author{Cheng-Feng Pu}
\affiliation{\institution{Tsinghua University}\country{China}}

\author{Ziyi Yang}
\affiliation{\institution{The University of Hong Kong}\country{China}}

\author{Yuan-Chen Guo}
\affiliation{\institution{VAST}\country{China}}

\author{Yan-Pei Cao}
\affiliation{\institution{VAST}\country{China}}
\authornote{Corresponding Author}

\author{Xiaojuan Qi}
\affiliation{\institution{The University of Hong Kong}\country{China}}
\authornotemark[2]

\makeatletter
\let\@authorsaddresses\@empty
\makeatother
\renewcommand{\shortauthors}{Yi-Hua Huang, et al.}

\begin{abstract}
Animatable 3D assets, defined as geometry equipped with an articulated skeleton and skinning weights, are fundamental to interactive graphics, embodied agents, and animation production. While recent 3D generative models can synthesize visually plausible shapes from images, the results are typically static. Obtaining usable rigs via post-hoc auto-rigging is brittle and often produces skeletons that are topologically inconsistent with the generated geometry. We present \textit{\name}, a unified framework that directly generates animate-ready 3D assets conditioned on a single image. Our key insight is to represent shape, skeleton, and skinning as mutually consistent \textit{$S^3$ Fields} (Shape, Skeleton, Skin) defined over a shared spatial domain. To enable the robust learning of these fields, we introduce two technical innovations: (i) a \textit{confidence-decaying skeleton field} that explicitly handles the geometric ambiguity of bone prediction at Voronoi boundaries, and (ii) a \textit{dual skin feature field} that decouples skinning weights from specific joint counts, allowing a fixed-architecture network to predict rigs of arbitrary complexity. Built upon a two-stage flow-matching pipeline, \name first synthesizes a sparse structural scaffold and then generates dense geometry and articulation in a structured latent space. Extensive experiments demonstrate that \name substantially outperforms state-of-the-art sequential baselines in rig validity and animation quality, generalizing effectively to in-the-wild images across diverse categories including animals, humanoids, and machinery. \textbf{Homepage:} \textcolor{magenta}{\url{https://yihua7.github.io/AniGen_web/}}
\end{abstract}

\begin{teaserfigure}
  \includegraphics[width=\textwidth]{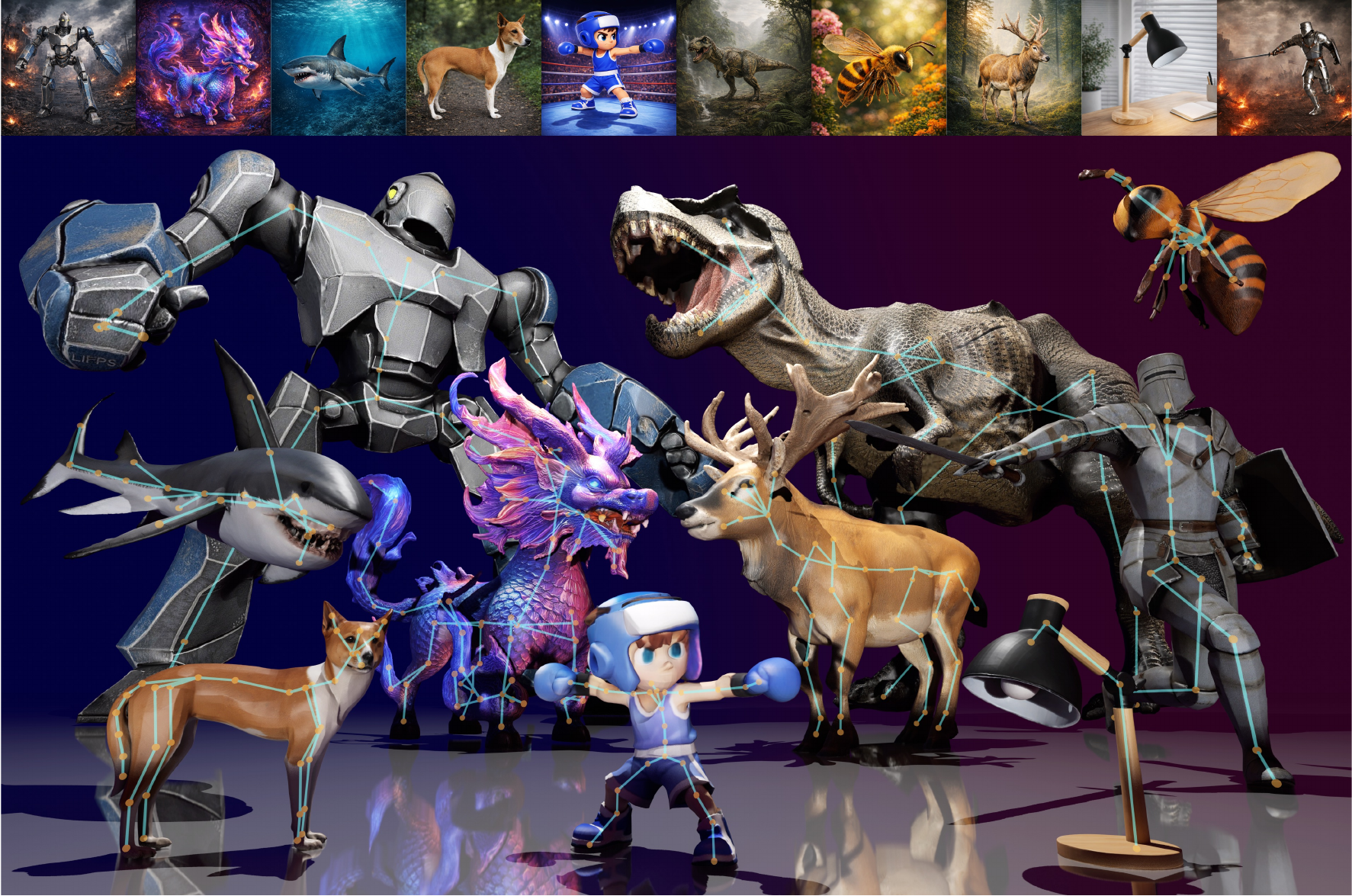}
  \caption{Given a single conditional image as input, \name generates a 3D shape along with its skeleton and skinning weights, supporting a wide range of 3D assets, including organic entities such as animals, cartoon characters, humans, and articulated man-made objects.}
  \label{fig:teaser}
\end{teaserfigure}

\maketitle

\section{Introduction}

The rapid advancement of generative 3D models~\citep{xiang2025structured,li2025triposg,zhang20233dshape2vecset,zhang2024clay,chen2025ultra3d,wu2024direct3d} has begun to democratize 3D content creation, enabling the synthesis of visually stunning geometry and appearance from simple text or image prompts. However, a significant gap remains between visual plausibility and functional utility. In interactive domains such as video games, virtual reality, and embodied AI, a 3D asset's utility depends on its \textit{animatability}. That is, it must be equipped with an articulated skeleton and precise skinning weights to enable posing, motion-capture driving, or physical simulation. Current generative models paradigms primarily yield static statues: models that process high-fidelity surface details but remain functionally inert, serving as mere decorations rather than interactive entities.

\begin{figure}[t]
    \begin{center}
        \includegraphics[width=\linewidth]{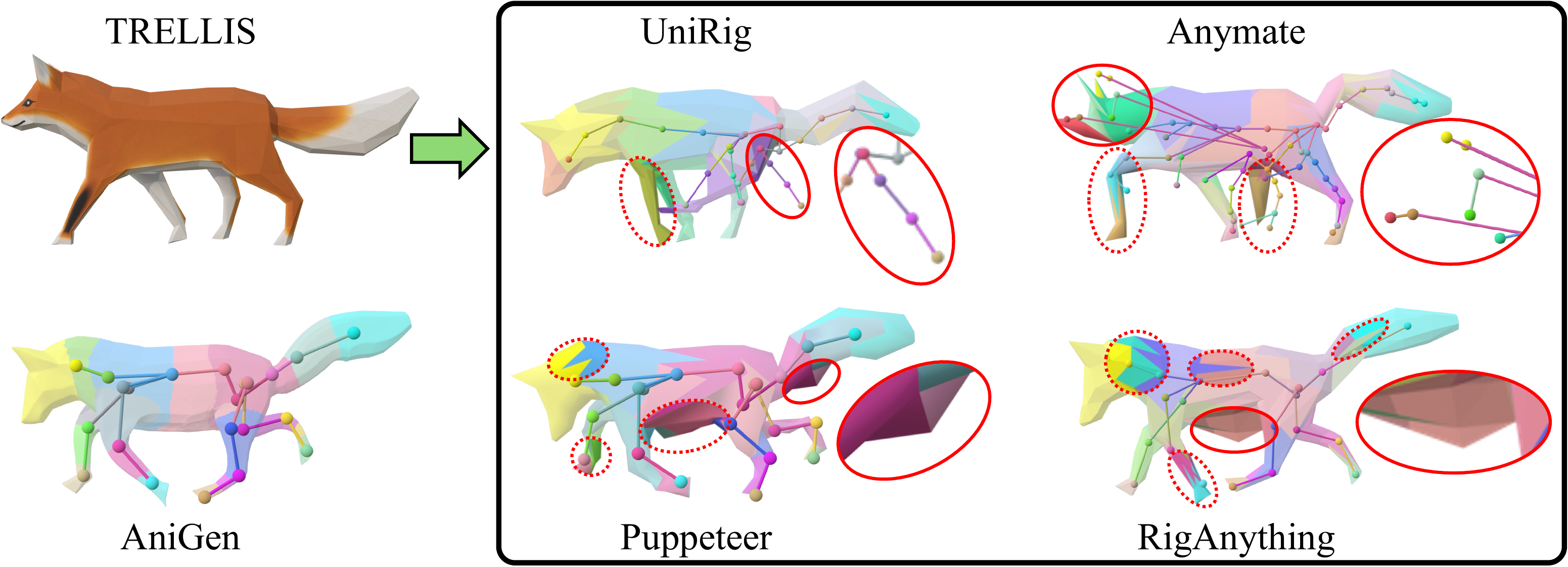}
    \end{center}
    \caption{We present a case study combining TRELLIS~\citep{xiang2025structured} with state-of-the-art auto-rigging baselines for animatable asset generation. Existing methods produce skeletons with missing bones (UniRig~\citep{zhang2025one}), unstructured rigs (Anymate~\citep{deng2025anymate}), or poor skinning (Puppeteer~\citep{songpuppeteer}, RigAnything~\citep{liu2025riganything}). The resulting animations expose practical failure modes of this pipeline, whereas \name generates plausible skeletons and skinning that support stable character animation.}
    \label{fig:case-study}
\end{figure}

A straightforward strategy to bridge this gap is a sequential ``generate-then-rig'' pipeline: first synthesize a static mesh using a state-of-the-art 3D generator~\citep{xiang2025structured}, followed by an off-the-shelf auto-rigging algorithm~\citep {zhang2025one,deng2025anymate,songpuppeteer,liu2025riganything}. Unfortunately, this decoupled approach is brittle. Unlike artist-authored assets, which are modeled with clean topology and articulation in mind, generated meshes often contain ``topological variances'', e.g., small geometric irregularities, diversely posed limbs, or ambiguous surface topology. While these variances may be visually negligible, they are catastrophic for auto-riggers, which rely on precise topological cues to infer kinematic chains. As shown in Figure~\ref{fig:case-study}, this mismatch often leads to anatomically incorrect skelections or skinning artifacts that cause unnatural shearing during animation.

In this work, we argue that this fragility stems from a fundamental \textit{representation mismatch}. Geometry and articulation are not disjoint attributes to be processed sequentially; rather, they are inherently entangled. The shape of a character is functionally determined by its underlying skeleton, and the validity of a skeleton is constrained by the volume of the shape. Therefore, treating rigging as a post-processing step ignores the intrinsic cross-modal priors that exist between shape and function. 

To address this, we present \textit{\name}, a unified generative framework that treats 3D shape, skeleton, and skinning as mutually consistent fields to be co-generated. Our key insight is to unify these distinct modalities into a common shared continuous representation, which we term \textit{$S^3$ Fields} ($S$hape, $S$keleton, $S$kin). By representing the skeleton not as a discrete graph but as a dense vector field, and skinning not as a sparse matrix but as a dual feature field, we enable all three components to share the same spatial domain and generative priors. Crucially, this unified representation enables a coherent compression-generation pipeline that can effectively ``grow'' the geometry and rig simultaneously. To make joint generation tractable, we introduce two key designs at the auto-encoding stage to compress the $S^3$ Fields into structured sparse latents~\citep{xiang2025structured}.
First, to address the inherent ambiguity of skeleton regression in regions equidistant to multiple bones (i.e., near Voronoi boundaries), we introduce a \textit{confidence-decaying skeleton field}. This mechanism explicitly down-weights ambiguous regions during auto-encoder training, focusing supervision on high-certainty areas near kinematic chains and thereby enforcing structural coherence.
Second, to accommodate the wide variation in joint counts across object categories, we propose a \textit{Dual Skin Field }representation, decoded via a pre-trained Skin Auto-Encoder (SkinAE). This design converts variable-cardinality skinning into a fixed-dimensional latent feature space, enabling consistent compression of rigs with arbitrary complexity.

Built upon these compressed latents, we train a sparse structured flow-matching model~\cite{xiang2025structured} to generate structured latent codes from image-conditioned noise, which are subsequently decoded into geometry and rigging simultaneously. \name demonstrates strong generalization capability, producing fully rigged, animate-ready 3D assets from single images across a broad range of categories, including humanoid characters, organic animals, and articulated machinery. Extensive experiments show that \name consistently outperforms sequential baselines in rig validity and animation quality, representing a practical step toward the scalable generation of functional 3D worlds.
\revise{We additionally show that this joint modeling of geometry and articulation does not compromise geometric fidelity relative to strong geometry-only generators, making the method practical rather than merely structurally correct.}

In summary, our contributions are:
\begin{itemize}
    \item \textit{Unified Generative Formulation:} We propose a holistic framework for the co-generation of shape, skeleton, and skinning as mutually aligned $S^3$ Fields, effectively eliminating the accumulation of errors inherent in sequential ``generate-then-rig'' pipelines.
    \item \textit{Confidence-Aware Skeleton Field:} We introduce a confidence-decaying vector field representation that resolves structural ambiguity, enabling robust graph extraction from noisy volumetric predictions.
    \item \textit{Joint-Count Agnostic Skinning:} We devise a Dual Skin Field and SkinAE training strategy that allows generative models to synthesize rigs of arbitrary complexity.
    \item \textit{State-of-the-Art Performance:} We demonstrate that \name produces high-fidelity, animate-ready assets from in-the-wild images, outperforming existing auto-rigging baselines.
\end{itemize}

\section{Related Work}

\subsection{Conditional 3D Generation}

Early text-to-3D generation methods are predominantly optimization-based. DreamField~\cite{jain2022zero} leverages CLIP~\cite{radford2021learning} to optimize NeRF~\cite{mildenhall2021nerf} renderings such that the reconstructed 3D scene aligns with a given text prompt. Building upon this paradigm, DreamFusion~\cite{poole2023dreamfusion} introduces Score Distillation Sampling (SDS), which employs a pretrained image diffusion model~\cite{ho2020denoising,songdenoising} to supervise NeRF optimization. Subsequent “dreamer” variants~\cite{wang2023prolificdreamer,yu2023text,liu2024syncdreamer,long2024wonder3d} further extend DreamFusion by improving the SDS formulation~\cite{wang2023prolificdreamer,yu2023text}, incorporating multi-view SDS supervision~\cite{liu2024syncdreamer,shimvdream}, and introducing normal-domain constraints~\cite{long2024wonder3d}. 

Despite their effectiveness, optimization-based approaches are computationally expensive, often requiring hours to generate a single asset. To address this limitation, feed-forward methods have been proposed to directly infer 3D representations. Methods such as Zero-1-to-3~\cite{liu2023zero}, Instant3D~\cite{li2024instant3d}, DMV3D~\cite{xu2024dmv3d}, and CAT3D~\cite{gao2024cat3d} first synthesize pose-conditioned multi-view images and then reconstruct 3D geometry using NeRF or 3D Gaussian Splatting (3DGS)~\cite{kerbl20233d}. In contrast, LRM~\cite{honglrm} formulates 3D reconstruction as a direct regression problem by employing a transformer~\cite{vaswani2017attention} to predict triplane representations from partial image observations. Building on this idea, TriplaneGaussian~\cite{zou2024triplane} and LGM~\cite{tang2024lgm} replace volumetric representations with 3DGS to improve rendering quality and geometric fidelity. TripoSR~\cite{TripoSR2024} further scales up LRM by leveraging larger datasets and refined architectural designs.

However, under partial observations, such as a single image or a small number of views, recovering a complete 3D shape is inherently a generative task rather than a pure reconstruction problem. Consequently, regression-based feed-forward methods tend to produce overly smoothed geometry, as they implicitly learn the data average and struggle to preserve fine-grained details. To mitigate this issue, VecSet~\cite{zhang20233dshape2vecset} proposes encoding 3D shapes into fixed-length token sequences and training diffusion models in the token space. Direct3D~\cite{wu2024direct3d} adopts VecSet tokens to predict triplanes that encode geometric structure, while Dora~\cite{chen2025dora} enhances detail preservation by adaptively sampling more points around sharp edges. Subsequent works, including TripoSG~\cite{li2025triposg} and Clay~\cite{zhang2024clay}, further improve performance by scaling up model capacity and leveraging larger, more carefully curated datasets.

More recently, TRELLIS~\cite{xiang2025structured} has emerged as a leading framework for high-quality 3D generation, demonstrating the effectiveness of a two-stage generation paradigm. In the first stage, TRELLIS generates a sparse voxel representation that captures the global structure of the shape. Conditioned on this sparse structure, the second stage refines geometry and details using rectified flow~\cite{liu2022flow} operating on sparse tensors. Several follow-up works improve upon this framework. SparseFlex~\cite{He_2025_ICCV} increases the voxel resolution in the first stage to $512^3$ to enhance geometric accuracy. Ultra3D~\cite{chen2025ultra3d} replaces the voxel-based first stage with VecSet representations and feeds voxelized outputs into the second-stage diffusion model. CUPID~\cite{huang2025cupid} improves image-conditioned generation by predicting per-voxel UV coordinates in the first stage, leading to better alignment with input images. Recently proposed TRELLIS 2~\cite{xiang2025native} further advances this paradigm by substantially increasing the first-stage resolution to $1536^3$ and incorporating high-quality PBR material prediction in the second stage, achieving state-of-the-art results and highlighting the advantages of two-stage 3D generation. Although these approaches produce geometry of exceptional quality, they are limited to static representations and lack the articulation capabilities essential for interactive 3D applications.

\begin{figure*}
    \begin{center}
        \vspace{-0mm}
        \includegraphics[width=1.0\linewidth]{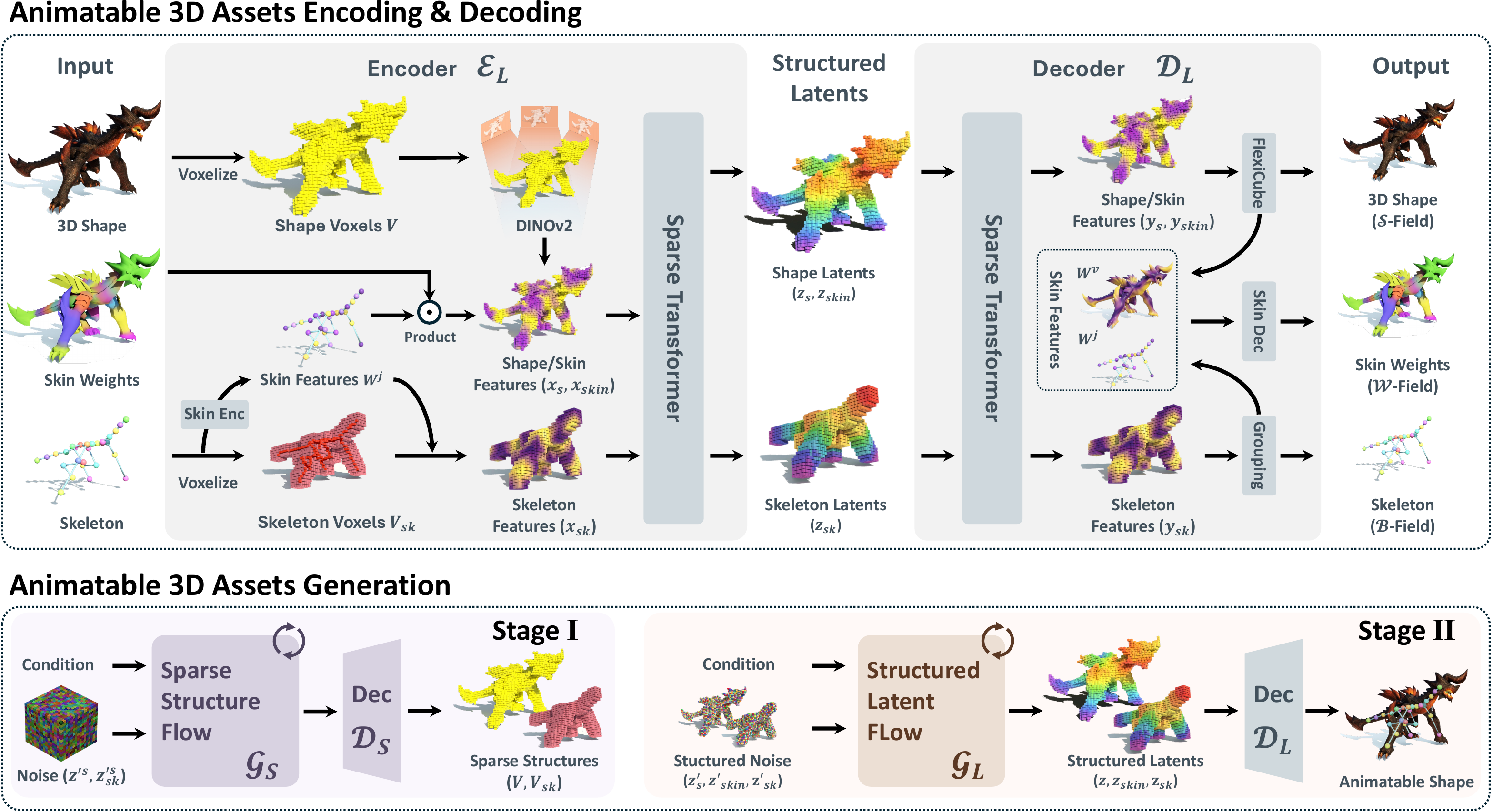}
    \end{center}
    \caption{Overview of the \name framework. The top panel illustrates the encoding and decoding of the $S^3$ fields using the structured latent autoencoder $\mathcal{E}_L$ and $\mathcal{D}_L$. From top to bottom, the three rows correspond to the shape, skin, and skeleton branches, respectively. The bottom panel depicts the generation pipeline for animatable 3D assets. Given an input image, the sparse structure flow model $\mathcal{G}_S$ predicts voxel scaffolds that serve as supports for the shape and skeleton. Conditioned on these supports, a structured latent flow $\mathcal{G}_L$ synthesizes the corresponding $S^3$ fields. Finally, the decoded outputs yield a 3D shape equipped with skeleton rigging and associated skinning, producing an animatable 3D asset.}
    \label{fig:pipeline}
\end{figure*}

\subsection{Automatic Rigging and Skinning}

\revise{Automatic rigging~\cite{Xu2019PredictingAS,RigMesh,Baerentzen14,FEQE} aims to equip static 3D meshes with skeletons and skinning weights to enable animation.} To address animation and rigging challenges, structural representations for 3D shapes were initially proposed by Marr et al.\cite{marr1978representation}, focusing on spatial organization rather than surface geometry. Pinocchio\cite{baran2007automatic} introduced a fully automated method for embedding template skeletons into arbitrary 3D meshes, enabling animation-ready rigs with minimal manual intervention. RigNet~\cite{xu2020rignet} further advanced rigging through deep learning but was limited to templates and T-poses. Subsequent works, such as Neural Blend Shapes~\cite{li2021learning} and TARig~\cite{ma2023tarig}, tackled rigging and skinning for humanoid characters, while MoRig~\cite{xu2022morig} used deforming point clouds for guidance. Other methods, such as NeuroSkinning~\cite{liu2019neuroskinning} and DRiVE~\cite{sun2025drive}, also focus on skeleton prediction and skinning for human characters.

More recently, general-purpose rigging methods have emerged. UniRig~\cite{zhang2025one} was one of the first to rig arbitrary shapes with an end-to-end auto-regressive model, followed by RigAnything~\cite{liu2025riganything}, which applied a joint diffusion model for sequential joint generation. Anymate~\cite{deng2025anymate} introduced a modular approach to predict joints, connectivity, and skinning weights, while Puppeteer~\cite{songpuppeteer} leveraged an auto-regressive transformer to predict skeletons and drive motion via optimization. However, these methods typically function as post-processing steps that depend heavily on the distribution of the input mesh. They often struggle with topological, shape-, and pose-variances common in generative outputs, motivating our end-to-end approach where geometry and rigging are co-generated.

\subsection{Generative Dynamic \& Articulated Assets}

Beyond generating static 3D content, researchers have also explored creating animatable or dynamic 3D objects. One direction involves 4D generation. DreamGaussian4D~\cite{ren2023dreamgaussian4d} extends optimization-based methods to dynamic generation. Cat4D~\cite{wu2025cat4d} and SVG~\cite{daisvg} use diffusion models to generate spatial-temporal frame matrices for dynamic scenes but are limited in view range. SMRNet~\cite{zhang2024skinned} and HMC~\cite{wang2023hmc} animate existing 3D shapes by retargeting them with reference animations, while SC4D~\cite{wu2024sc4d} and DeformSplat~\cite{kim2025rigidity} optimize 3D shapes into dynamic motion with sparse control and local rigidity~\cite{huang2024sc}. AnimateAnyMesh~\cite{Wu_2025_ICCV} generates deformed mesh sequences using a feedforward model, and SS4D~\cite{li2025ss4d} predicts sequential sparse structures and latent representations to produce dynamic scenes. While these methods generate dynamic content, they often lack flexible controllers to freely adjust pose and motion.

Another direction explores combining rigging with motion synthesis or part-level articulation. Anytop~\cite{gat2025anytop} animates skeletons based on semantic joint names using a transformer with skeletal and temporal attention. AnimaX~\cite{huang2025animax} generates multi-view videos of animated characters and derives pose parameters via triangulation and inverse kinematics. Similarly, AnimaMimic~\cite{xie2025animamimic} combines UniRig with video priors to drive motion. Make-it-Poseable~\cite{guo2025make} bypasses skinning by directly animating meshes with their associated skeletons. For articulated objects, ArtiLatent~\cite{chen2025artilatent} generates sparse structures with articulation labels (e.g., part types and joint types), enabling dynamic objects like drawers or cabinets. Particulate~\cite{li2025particulate} applies a part articulation transformer to predict articulation labels, supporting downstream embodied AI tasks. Unlike these works, which often focus on rigid parts or video-driven priors, our framework unifies shape, skeleton, and skinning generation to produce fully animatable organic assets.

\section{Method}

\subsection{Overview}
\label{sec:overview}
Our goal is to generate a fully functional, animatable 3D asset from a single RGB image $\mathbf{I}$. Formally, an animatable asset is a tuple $\mathcal{A} = (\mathcal{M}, \mathcal{K}, \mathcal{W})$, consisting of a 3D mesh geometry $\mathcal{M}$, a hierarchical skeleton $\mathcal{K}$ (joints and connectivity), and skinning weights $\mathcal{W}$ that bind the geometry to the skeleton.

Existing approaches typically treat this as a sequential pipeline: generating a static mesh $\mathcal{M}$ first, and then predicting $\mathcal{K}$ and $\mathcal{W}$ via post-processing. This separation often leads to topological mismatches, where the generated geometry lacks the structural integrity required for articulation (e.g., fused limbs).

\name fundamentally departs from this paradigm by modeling $S$hape, $S$keleton, and $S$kinning as a unified, spatially aligned representation which we term {$S^3$ Fields} (Sec.~\ref{sec:s3-field}). By representing articulation properties as continuous fields sharing the same spatial domain as the geometry, we ensure mutual consistency.

To synthesize these fields from a single image, we devise a two-stage generative pipeline (see Fig.~\ref{fig:pipeline}):
\begin{itemize}
    \item \textit{Representation \& Compression:} We first learn to compress the continuous $S^3$ Fields into a compact, sparse structural representation. We employ a sparse structure auto-encoder ($\mathcal{E}_S, \mathcal{D}_S$) to capture the coarse spatial layout and a structured latent auto-encoder ($\mathcal{E}_L, \mathcal{D}_L$) to encode fine-grained geometry and skinning features into a low-dimensional latent space (Sec.~\ref{sec:dae}).
    \item \textit{Generative Flow:} We then train two flow-matching models to synthesize these representations from the input image. A sparse structure flow model first predicts the spatial occupancy and skeleton graph, followed by a structured latent flow model that ``paints'' the geometry and skinning details onto this structure (Sec.~\ref{sec:flow-model}).
\end{itemize}

\subsection{$S^3$ Fields}
\label{sec:s3-field}

A core challenge in generating animatable assets is that skeletons and skinning weights are traditionally irregular data structures (graphs and sparse matrices), which are difficult to generate using standard convolutional or transformer architectures that assume fixed grid topologies. We resolve this by lifting them into continuous fields defined over the shared 3D domain $\mathbb{R}^3$. We collectively term this representation $S^3$ Fields.

\subsubsection{Shape Field $\mathcal{S}$} 
\label{sec:shape-field}
Following recent state-of-the-art 3D generation method~\cite{xiang2025structured}, 
we define the shape field over a set of active sparse voxels $\mathcal{V} \subset \{0, 1\}^{N^3}$ that track the mesh surface. Formally, the Shape Field $\mathcal{S}$ maps each active voxel coordinate $v \in \mathcal{V}$ to a set of local geometric and appearance parameters required by FlexiCubes~\citep{shen2023flexible} for mesh extraction:
\begin{equation}
    \mathcal{S}(v) = \left[ \mathbf{d}, \mathbf{n}, \mathbf{c}, \mathbf{f}_{\text{flex}} \right] \in \mathbb{R}^{C_{\mathcal{S}}},
\end{equation}
where $\mathbf{d} \in \mathbb{R}^8$ denotes the signed distances at the voxel corners, $\mathbf{n} \in \mathbb{R}^{8 \times 3}$ and $\mathbf{c} \in \mathbb{R}^{8 \times 3}$ represent corner normals and colors, and $\mathbf{f}_{flex}$ contains the FlexiCubes-specific interpolation and splitting weights~\citep{shen2023flexible}. This explicit parameterization enables the extraction of high-quality, watertight meshes $\mathcal{M}$ directly from the field components.
\revise{Since color is stored in the same field, the final textured mesh can be obtained during surface extraction.}

\subsubsection{Skeleton Field  $\mathcal{B}$} 
\label{sec:skeleton-field}
A skeleton is conventionally represented as a set of joints organized in a tree structure in Euclidean space. However, such a discrete graph representation is ill-suited for generative modeling: it lacks a fixed spatial support, does not align naturally with volumetric or grid-based shape representations, and is difficult to co-generate with geometry using shared model structures.
To this end, we represent the skeleton $\mathcal{K}$ not as a graph, but as a dense vector field, allowing it to be co-generated with the shape using the same model architecture. Let $\mathcal{K} = \{J_1, \dots, J_M\}$ be the set of discrete joints in Euclidean space. The Skeleton Field $\mathcal{B}: \mathbb{R}^3 \to \mathbb{R}^6$ maps any point $x \in \mathbb{R}^3$ to a vector pair pointing to its nearest joint $j(x) \in \mathcal{K}$ and that joint's parent $p(x) \in \mathcal{K}$:
\begin{equation}
    \mathcal{B}(x) = \left[ (j(x) - x) \oplus (p(x) - x) \right],
\end{equation}
where $\oplus$ denotes concatenation. This relative parameterization ensures the field is translation-invariant and local.

\begin{figure}[h]
    \begin{center}
        \includegraphics[width=\linewidth]{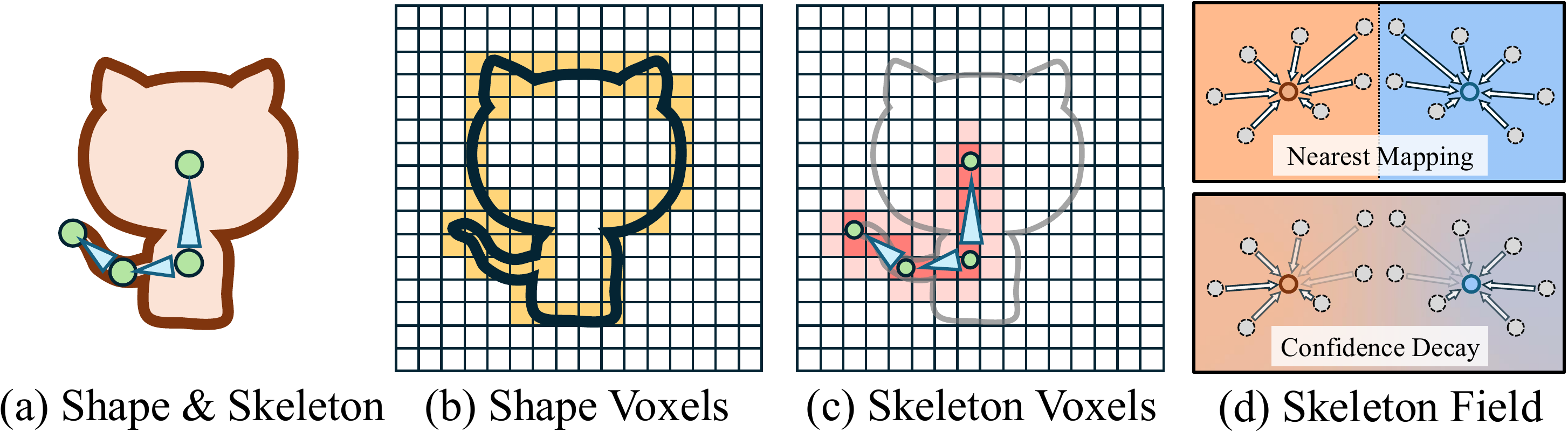}
    \end{center}
    \caption{Illustration of the (b) shape voxels $\mathcal{V}$, (c) skeleton voxels $\mathcal{V}_{sk}$, and the (d) confidence-decaying nearest-neighbor joint field.}
    \label{fig:skeleton-voxels}
\end{figure}

\paragraph{Voxel Support $\mathcal{V}_{sk}$}
To model this field efficiently, we parameterize it using latent features distributed across a specific set of sparse voxels $\mathcal{V}_{sk}$. Critically, we cannot simply reuse the shape voxels $\mathcal{V}$ for this purpose, as $\mathcal{V}$ tracks the surface boundary, whereas skeleton joints often reside deep within the object's volume (far from surface voxels). Therefore, we define $\mathcal{V}_{sk}$ as the set of voxels intersected by the skeleton bones. To ensure robust coverage and prevent disconnection during generation, we dilate $\mathcal{V}_{sk}$ with a radius of $2$ voxels. This ensures the field is defined on a continuous, volumetric structure enveloping the kinematic chains (see Fig.~\ref{fig:skeleton-voxels}).

\paragraph{Confidence-Aware Prediction} 
A major challenge in regression-based skeleton prediction arises near Voronoi boundaries between joints, where the identity of the nearest bone changes abruptly. At these locations, the regression target becomes discontinuous, causing ambiguous supervision and unstable gradients during training. Standard Bayesian uncertainty learning strategies~\cite{kendall2017uncertainties} often struggle here, as they rely on a learned variance that is difficult to weight against the regression loss and cannot guarantee low confidence scores in ambiguous regions (see Sec.~\ref{sec:ablation}).
To mitigate this, we augment the field with a scalar \textit{confidence score} $c(x) \in [0, 1]$. We explicitly supervise the confidence using a geometric metric that reflects the ambiguity of joint assignment. For a voxel center $v_c$, let $j^{gt}$ be the nearest joint and $j_{2nd}^{gt}$ be the second nearest. We define the ground truth (GT) confidence as:
\begin{equation}
\label{eq:confidence}
    c_{gt}(v_c) = 1 - \frac{\|v_c - j^{gt}\|^2}{\|v_c - j_{2nd}^{gt}\|^2} \in [0, 1].
\end{equation}

The model predicts an extra scalar field to fit this target. Crucially, we use $c_{gt}$ to weight the regression loss for joint and parent predictions. This forces the model to focus on the high-certainty regions near bones while suppressing gradients from ambiguous boundaries. Additionally, the learned confidence field facilitates the next joint recovery stage.

\begin{figure}[h]
    \begin{center}
        \includegraphics[width=\linewidth]{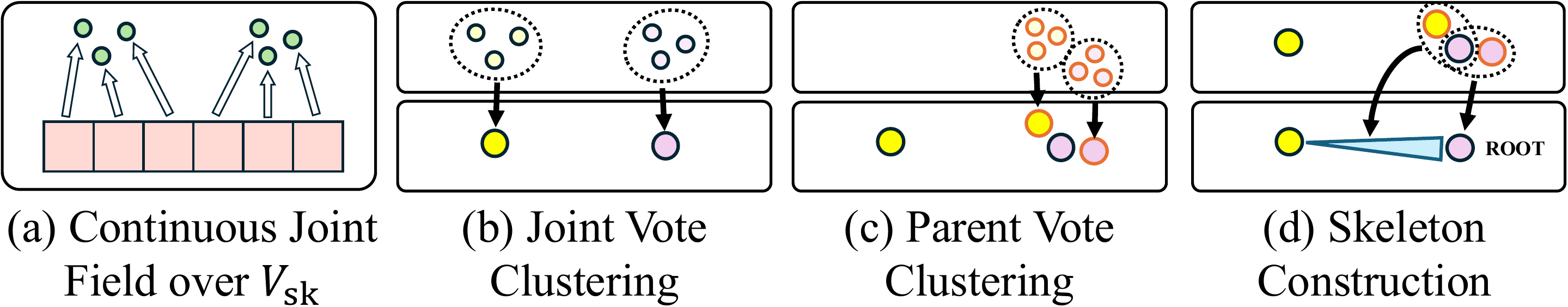}
    \end{center}
    \caption{\revise{Illustration of the conversion from a continuous field to a discrete skeleton. The voting results of the joint field in (a) are clustered to obtain discrete joints in (b) and parent nodes in (c). The skeletal topology is then determined by assigning each parent to its nearest joint as in (d).}}
    \label{fig:field-to-skeleton}
\end{figure}

\paragraph{From Continuous Field to Discrete Skeleton}
During inference, we can recover the discrete skeleton $\mathcal{K}$ from the predicted field $\mathcal{B}$ and confidence $c$.
For every center $v_i \in \mathcal{V}_{sk}$, we derive a voting joint $\hat{j}_i$ and corresponding parent $\hat{p}_i$ using the predicted Skeleton Field $\hat{\mathcal{B}}(v)$. We then employ a confidence-weighted iterative clustering algorithm (detailed in Alg.~\ref{alg:meanshift_grouping}) to enhance the joint prediction accuracy: we iteratively shift the voting points toward the centroid of their local neighborhood, weighted by their predicted confidence. This effectively concentrates votes into tight clusters. The final confidence-weighted centroids are identified as joints, and isolated low-confidence points are filtered out to remove noise. Additionally, the clustering process estimates the parent location for each cluster. By connecting the cluster centroid to its closest estimated parent, the joints are sequentially linked to form the skeleton. \revise{Fig.~\ref{fig:field-to-skeleton} provides an intuitive illustration of this conversion.}

\subsubsection{Dual Skin Field  $\mathcal{W}$} 
\label{sec:skin-field}
Skinning weights represent a normalized probability distribution indicating how each vertex follows the motion of the underlying skeleton. 
Parameterizing this relationship directly with a neural network is non-trivial because the number of joints, $N_j$, varies significantly across different asset categories. This variability precludes the use of standard regression layers with fixed output dimensions.

To overcome this, we propose a \textit{Dual Skin Field} formulation that factorizes the skinning function into two implicit feature fields defined over the shape voxels $\mathcal{V}$ and skeleton voxels $\mathcal{V}_{sk}$, respectively:
\begin{equation}
    \mathcal{W}, \mathcal{W}_{sk}: \mathbb{R}^3 \rightarrow \mathbb{R}^{D_{\text{skin}}},
\end{equation}
where $D_{skin}$ denotes the dimension of the shared latent embedding space.
\begin{itemize}
    \item The \textit{Surface Skin Field} $\mathcal{W}$ encodes local deformability and segmentation cues of the geometry. This field is instantiated as $y_{\text{skin}}$ in the auto-encoder, as illustrated in Fig.~\ref{fig:pipeline}.
    \item The \textit{Skeleton Skin Field} $\mathcal{W}_{sk}$ encodes the influence characteristics of the skeletal structure. This field is instantiated as parts of $y_{sk}$ in the auto-encoder, as illustrated in Fig.~\ref{fig:pipeline}.
\end{itemize}
This representation effectively decouples the network architecture from the specific joint count. To recover the explicit skinning weights $w_v \in \mathbb{R}^{N_j}$ for a query vertex $v$ and a generated set of joints $\{ j_i \}_{i=1}^{N_j}$, we query the fields to obtain the vertex feature $\mathcal{W}(v)$ and the joint features $\{ \mathcal{W}_{sk}(j_i) \}_{i=1}^{N_j}$. These features are mapped to a shared compatibility space via a lightweight MLP, and the final weights are computed using a cross-attention operation, ensuring that $w_v$ forms a valid partition of unity (sums to 1). For more implementation details, please refer to Sec.\ref{sec:skinae} and Fig.\ref{fig:skin-ae}. This allows our generative model to predict fixed-size feature maps that naturally generalize to rigs of arbitrary complexity.

\SetAlgoNlRelativeSize{-1}
\SetAlgoNlRelativeSize{-2}
\begin{algorithm}[h]
\caption{Field-to-Skeleton Clustering}
\label{alg:meanshift_grouping}
\DontPrintSemicolon
\SetKwInOut{KwIn}{Input}
\SetKwInOut{KwOut}{Output}

\KwIn{$\mathbf{J},\mathbf{P}\in\mathbb{R}^{N\times 3}$; confidences $\mathbf{c}\in\mathbb{R}^{N\times 1}$ (default ones);
threshold $\tau$; bandwidth $h$; min cluster size $s_{\min}$.}
\KwOut{Grouped joints $\bar{\mathbf{J}}\in\mathbb{R}^{M\times 3}$; parent indices $\boldsymbol{\pi}\in\{-1,\dots,M-1\}^{M}$.}
$\mathbf{S}\leftarrow \mathbf{J}$\;
\For{$t\leftarrow 1$ \KwTo 10}{
  \For{$i\leftarrow 1$ \KwTo $|\mathbf{S}|$}{
    $\mathcal{N}_i\leftarrow$ neighbors of $\mathbf{S}_i$ within radius $h$\;
    $w_{ij}\leftarrow c^J_j \exp(-\frac{\|\mathbf{S}_j-\mathbf{S}_i\|^2}{2h^2})\ \ \forall j\in\mathcal{N}_i$\;
    $\mathbf{S}'_i \leftarrow \dfrac{\sum_{j\in\mathcal{N}_i} w_{ij}\mathbf{S}_j}{\sum_{j\in\mathcal{N}_i} w_{ij}+10^{-8}}$\;
  }
  \If{$\max_i\|\mathbf{S}'_i-\mathbf{S}_i\|\le \frac{1}{10}h$}{\textbf{break}\;}
  $\mathbf{S}\leftarrow \mathbf{S}'$\;
}
Cluster $\mathbf{S}$ with merge radius $r=\frac{h}{2}$ to get labels $\ell_i\in\{1,\dots,N_c\}$\;
\For{$k\leftarrow 1$ \KwTo $N_c$}{
  $\tilde{\mathbf{J}}_k \leftarrow \dfrac{\sum_{i:\ell_i=k} c^J_i\mathbf{J}_i}{\sum_{i:\ell_i=k} c^J_i}$;
  $\tilde{\mathbf{P}}_k \leftarrow \dfrac{\sum_{i:\ell_i=k} c^P_i\mathbf{P}_i}{\sum_{i:\ell_i=k} c^P_i}$\;
}
Keep clusters with $|\{i:\ell_i=k\}| \ge s_{\min}$ to obtain $\bar{\mathbf{J}},\bar{\mathbf{P}}$\;
\For{$k\leftarrow 1$ \KwTo $M$}{
  $\pi_k \leftarrow \arg\min_u \|\bar{\mathbf{P}}_k-\bar{\mathbf{J}}_u\|_2$\;
}
\Return $\bar{\mathbf{J}},\boldsymbol{\pi}$\;
\end{algorithm}

\subsection{Latent Representation of $S^3$ Fields}
\label{sec:dae}

To enable scalable generative modeling of complex $S^3$ fields, we must project them onto a compact, lower-dimensional manifold. We achieve this via a hierarchical compression strategy that disentangles coarse structural topology from fine-grained geometry and articulation. Unlike standard Variational Auto-Encoders (VAEs)~\citep{kingma2013auto} commonly used in latent diffusion, we employ \textit{Denoising Auto-Encoders (DAEs)}. This design choice, aligned with recent findings in generative modeling~\citep{yang2025latent,yao2025towards}, provides a latent space that is structurally better suited for the linear interpolation trajectory of flow matching.

To ensure training stability and prevent the model from exploiting unbounded feature magnitudes to trivialize the reconstruction task (which leads to singularities in the flow field), we strictly normalize all latent features onto a unit $\ell_1$-norm hypersphere. \revise{Without this constraint, we observe that the DAE can artificially inflate the latent norm to enlarge pairwise sample distances, which improves reconstruction shortcuts but makes the subsequent flow-matching objective substantially harder.} During training, we perturb the clean encoded latent $z$ with standard Gaussian noise $n$, mixed via a coefficient $t \in [0, t_{\max}]$, \revise{i.e., $z_{\text{sample}} = t \cdot n + (1-t)\cdot z$.} To further stabilize the learning process, we implement a curriculum schedule where $t_{\max}$ is linearly increased from $0$ to $0.75$ over the course of training.

\subsubsection{Prerequisite: Skin Auto-Encoder (SkinAE)}
\label{sec:skinae}
Before introducing the main auto-encoder that compresses the full $S^3$ Fields, we first describe how skinning information is encoded into latent features. Skinning weights $W \in \mathbb{R}^{N_v \times N_j}$ inherently depend on the number of joints $N_j$, which varies significantly across asset categories (e.g., $N_j = 10$ for a fish versus $N_j = 52$ for a humanoid). This variable cardinality makes explicit skinning matrices incompatible with standard neural networks that require fixed input channel dimensions.

To address this issue, we introduce \textit{SkinAE}, which learns a \textit{joint-count--agnostic} representation of skinning. Rather than representing skinning as a joint-indexed matrix, SkinAE factorizes it into fixed-dimensional \textit{joint embeddings} and \textit{vertex embeddings}, thereby decoupling the representation from the number of joints while preserving articulation structure.

\begin{figure}[h]
    \begin{center}
        \vspace{-0mm}   \includegraphics[width=1.0\linewidth]{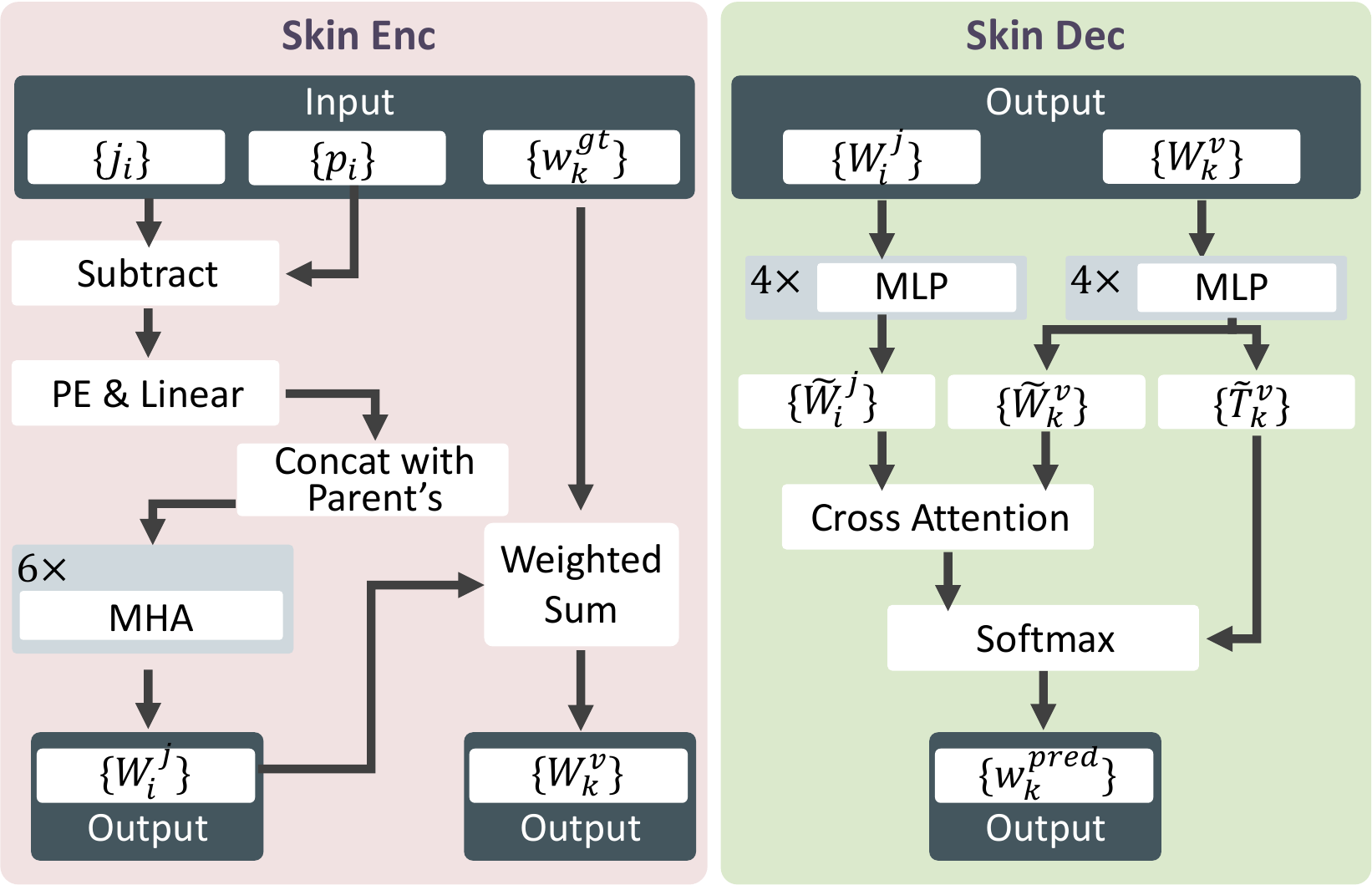}
    \end{center}
    \caption{
    SkinAE encodes skeleton joints ${j_i}$ and parents ${p_i}$ into joint skin features ${W^j_i}$, which are averaged by GT skin weights ${w_k^{gt}}$ to obtain vertex skin features ${W^v_k}$. The decoder processes ${W^j_i}$ and ${W^v_k}$ via MLPs, producing ${\tilde{W}_i^j}$, ${\tilde{W}_v^k}$, and vertex temperatures ${\tilde{T}_v^k}$. Final skin weights are obtained through cross-attention and Softmax with temperature adjustment.
    }
    \label{fig:skin-ae}
\end{figure}

\paragraph{Architecture.}
SkinAE comprises a Transformer-based encoder and a lightweight MLP decoder, as shown in Fig.~\ref{fig:skin-ae}.
\begin{enumerate}
    \item \textit{Joint Encoding:} Given a set of skeleton joints ${j_i}_{i=1}^{N_j}$, we compute parent-relative edge vectors $e_i = j_i - p_i$, apply positional encoding (PE), and concatenate them with parent features to incorporate structural information. A Transformer encoder processes these to generate joint skin features ${W_i^j \in \mathbb{R}^C}, (C=4)$.
    \item \textit{Vertex Encoding:} Corresponding vertex embeddings $\{W_k^v \in \mathbb{R}^C\}$ are computed via the \textit{skin-weighted average} of the joint embeddings. It makes the vertex skin features of a fixed channel dimension independent of $N_j$. This compresses the sparse skinning matrix into a dense and fixed-channel vertex field.
    \item \textit{Decoding:} A lightweight decoder MLP lifts these features to a higher dimension ($\mathbb{R}^{64}$) and reconstructs the original skinning weights via \textit{channel-wise compatibility}:
    \begin{equation}
        w_{k}(v) = \mathrm{Softmax}_i \left( \frac{1}{\tilde{T}_k^v} \langle \tilde{W}_k^v, \tilde{W}_i^j \rangle \right).
    \end{equation}
\end{enumerate}
Here, $\tilde{W}$ denotes output features produced by the decoder MLPs (see Fig.~\ref{fig:skin-ae}). We pre-train SkinAE using 3D points sampled directly from the mesh surface and freeze the network weights for the subsequent stages. Empirically, we find this pre-training strategy is essential for the convergence and performance of the structured latent auto-encoder (as demonstrated in Sec.~\ref{sec:ablation}). This step effectively converts the variable-cardinality skinning regression task into a \textit{fixed-channel feature matching task}, enabling the generative models to synthesize skinning fields for rigs of any complexity.

\subsubsection{Sparse Structure Auto-Encoder $\mathcal{E}_S~\&~\mathcal{D}_S$}
The first stage of our pipeline captures the coarse spatial layout. The Sparse Structure Auto-Encoder learns compact discrete representations from two aligned volumetric inputs: the shape occupancy grid $\mathcal{V} \in \{0, 1\}^{64^3}$ and the skeleton occupancy grid $\mathcal{V}_{sk}$ (defined in Sec.~\ref{sec:skeleton-field}).

The encoder $\mathcal{E}_S$ processes each input with a lightweight 3D convolutional network. 
We utilize strided 3D convolutions to downsample the spatial resolution, followed by multi-resolution 3D residual blocks to capture multi-scale structural dependencies. This process yields two distinct compressed latent volumes: $z^{s} \in \mathbb{R}^{16^3 \times 8}$ for the shape structure and $z^s_{sk} \in \mathbb{R}^{16^3 \times 4}$ for the skeleton structure.
Conversely, the decoder $\mathcal{D}_S$ maps these latents back to the original resolution using progressive 3D upsampling and residual blocks. The network terminates in two separate output heads that reconstruct the binary occupancy probabilities for the shape and skeleton volumes, respectively.

\subsubsection{Structured Latent Auto-Encoder $\mathcal{E}_L~\&~\mathcal{D}_L$} \label{sec:structured-ae}
The Structured Latent Auto-Encoder ($\mathcal{E}_L, \mathcal{D}_L$) serves as the core engine for high-fidelity reconstruction (see Fig.~\ref{fig:pipeline}, top). Using the sparse voxels $\mathcal{V}$ and $\mathcal{V}_{sk}$ as scaffolds, it encodes the fine-grained $S^3$ fields into a continuous latent space.

\paragraph{Input Feature Construction}
We construct voxel-aligned sparse inputs from both multi-view and geometry observations:
\begin{itemize}
    \item \textit{Shape ($x_\mathrm{s}$):} Defined on $\mathcal{V}$. We back-project multi-view DINOv2~\citep{oquab2024dinov2} features onto the voxel grid following TRELLIS~\citep{xiang2025structured}.
    \item \textit{Skinning ($x_\mathrm{skin}$):} Defined on $\mathcal{V}$. For each occupied voxel, we query the nearest point on the ground-truth mesh and assign it the corresponding vertex embedding $W^v$ derived from the frozen SkinAE.
    \item \textit{Skeleton ($x_\mathrm{sk}$):} Defined on $\mathcal{V}_{sk}$. We concatenate the positional embeddings of the nearest joint and its parent, along with the joint embedding $W^j$ from SkinAE.
\end{itemize}

\paragraph{Encoding \& Decoding}
The encoder $\mathcal{E}_L$ employs a multi-stream sparse Transformer backbone (12 Swin-style blocks~\citep{liu2021swin}) to process these inputs, producing three latent volumes $(z_s, z_{\mathrm{skin}}, z_{\mathrm{sk}})$ with channel dimensions $8$, $4$, and $4$, respectively.
The decoder $\mathcal{D}_L$ applies the same multi-stream sparse Transformer backbone to obtain hidden features $(h_s, h_{\mathrm{skin}}, h_{\mathrm{sk}})$. It then upsamples the shape and skin latents to high resolution ($256^3$) to predict: (i) the Shape Field $y_s$ for FlexiCubes extraction, and (ii) the Vertex Feature Field $y_{skin}$ for predicting $\{W_k^v\}$. Simultaneously, the skeleton branch decodes the Skeleton Field $y_\mathrm{sk}$ (vectors and confidence) and Joint Feature Field for predicting $\{W_i^j\}$.
The final asset is assembled by extracting the mesh via Marching FlexiCubes, clustering the skeleton field into discrete joints (Sec.~\ref{sec:skeleton-field}), and decoding the skinning weights using the SkinAE decoder on the predicted features. The reconstructed shape is thus rigged with a predicted skeleton, as illustrated in Fig.~\ref{fig:pipeline}.

\paragraph{Confidence-Weighted Supervision}
We supervise the geometry with standard rendering losses (depth, normal, color). For the skeleton and skeleton-side skinning fields ($\mathcal{W}_{sk}$), which are both represented by $y_{\mathrm{sk}}$, we employ confidence-weighted supervision to mitigate border ambiguity (see Sec.~\ref{sec:skeleton-field}). The regression loss for a joint prediction $j_{v_c}$ at voxel $v_c$ is:
\begin{equation}
L_J=\mathbb{E}_{v_c}\!\left[c_{gt}(v_c)\,\lVert j_{v_c}-j_{v_c}^{gt}\rVert_2^2\right],
\end{equation}
where $c_{gt}$ is the ground-truth confidence (Eq.~\ref{eq:confidence}). The same weighting is applied to parent predictions and skin feature $W^j$ predictions in the skeleton branch.
\revise{Furthermore, the structured latent auto-encoder is trained with feature reconstruction losses for the surface-side skinning embeddings. The sparse-structure auto-encoder, augmented with an additional branch to encode $\mathcal{V}{sk}$, is optimized using binary occupancy reconstruction losses on both $\mathcal{V}$ and $\mathcal{V}{sk}$.}

\begin{figure}[h]
    \begin{center}
        \vspace{-0mm}
        \includegraphics[width=1.0\linewidth]{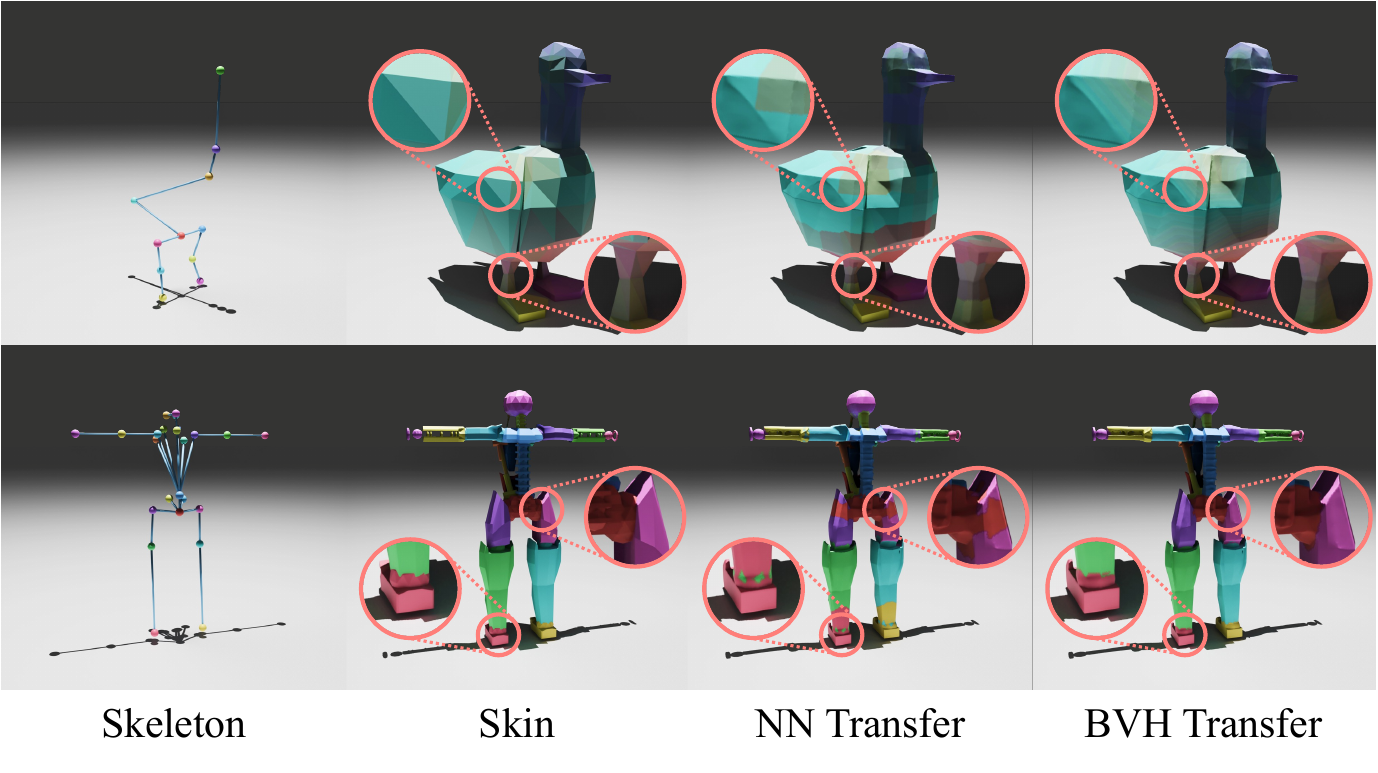}
    \end{center}
    \caption{BVH-based skin transfer vs. nearest-vertex (NN) transfer. We implement a CUDA BVH supporting multi-device deployment and save/restore.}
    \label{fig:skin_transfer}
\end{figure}

\paragraph{BVH-Accelerated Skin Transfer.}
To supervise the predicted vertex skin features $\{W_k^v\}$, we must match them to the ground truth (GT). Since the predicted mesh topology differs from the GT mesh, we transfer GT skin features via nearest-surface interpolation. To make this efficient during training, we implement a CUDA-based Bounding Volume Hierarchy (BVH) that caches the GT geometry. This reduces average query time from 48.6\,ms to 2.6\,ms ($18.6\times$ speedup). Crucially, as shown in Fig.~\ref{fig:skin_transfer}, BVH-based barycentric transfer is more robust to uneven vertex sampling than simple nearest-vertex matching.

\subsection{Generative Flow Model}
\label{sec:flow-model}
\begin{figure}[h]
    \begin{center}
        \vspace{-0mm}
        \includegraphics[width=1.0\linewidth]{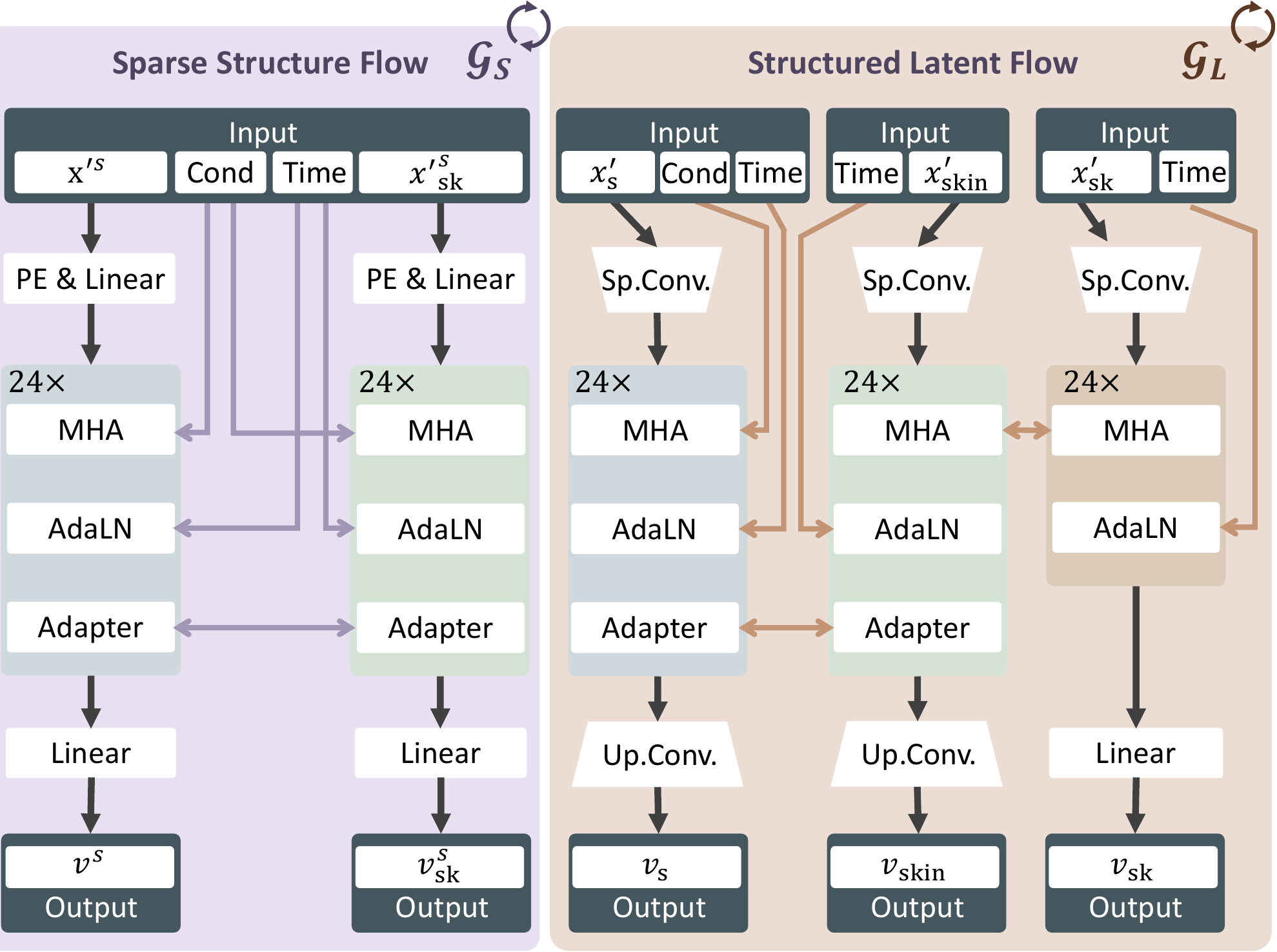}
    \end{center}
    \caption{The architectures of the AniGen flow models, $\mathcal{G}S$ and $\mathcal{G}L$. The input to $\mathcal{G}S$ comprises noisy volumetric features encoding geometry, $z'^s$, and skeleton information, $z'^s{\mathrm{skl}}$. $\mathcal{G}L$ processes noisy structured latents representing geometry, $z'{\mathrm{s}}$, skin, $z'{\mathrm{skin}}$, and skeleton, $z'{\mathrm{skl}}$. These flow models predict the velocities of the noisy features and iteratively denoise them using the Euler method.} 
    \label{fig:flow_model}
\end{figure}

We model the generation of animatable assets as a conditional flow matching problem. Formally, we aim to learn a velocity field $v_t$ that transports a standard Gaussian distribution $\pi_0 = \mathcal{N}(0, I)$ to the data distribution $\pi_1$ of our compressed representations.
To handle the complex interplay between topological structure and dense attributes (i.e., geometry and skinning weights), we decompose the generation process into two cascaded stages: \textit{Sparse Structure Flow} ($\mathcal{G}_S$) and \textit{Structured Latent Flow} ($\mathcal{G}_L$). Fig.~\ref{fig:flow_model} illustrates our flow model architecture.

\subsubsection{Stage I: Sparse Structure Flow $\mathcal{G}_S$}
The first stage constructs the \textit{scaffold} of the asset. Conditioned on image features, $\mathcal{G}_S$ predicts the active sparse voxel sets for both the shape ($\mathcal{V}$) and the skeleton ($\mathcal{V}_{sk}$).
We instantiate $\mathcal{G}_S$ as a dual-stream Transformer. Rather than concatenating shape and skeleton into a single sequence (which obscures their distinct topological roles), we process them in parallel branches:
\begin{itemize}
    \item \textit{Shape Branch:} Predicts the binary occupancy of surface-crossing voxels.
    \item \textit{Skeleton Branch:} Predicts the binary occupancy of bone-containing voxels.
\end{itemize}
\paragraph{Cross-Structural Adapters.} A naïve dual-stream approach risks generating a skeleton that drifts outside the mesh. To enforce spatial compatibility, we introduce lightweight linear adapters that exchange information between the two branches at every Transformer block. This bidirectional fusion ensures the skeleton ``grows'' strictly within the geometric bounds of the shape.

\subsubsection{Stage II: Structured Latent Flow $\mathcal{G}_L$}
\label{sec:slat-flow}
Given the generated scaffolds, the second stage synthesizes the $S^3$ field latent features. $\mathcal{G}_L$ is trained to denoise the latent codes $z_{\mathrm{s}}$, $z_{\mathrm{skin}}$, and $z_{\mathrm{sk}}$.

\paragraph{Architecture}
Similar to Stage I, we employ a multi-branch architecture. Since geometry and skinning share the same spatial domain ($\mathcal{V}$), we process them in a primary branch, while the skeleton latent $z_{\mathrm{sk}}$ (defined on $\mathcal{V}_{sk}$) is processed in a secondary branch. We again utilize adapter layers to enforce consistency between the predicted skinning features and the underlying skeletal structure.

\paragraph{Controllable Joint Density}
A key advantage of our joint-count agnostic representation (Sec.~\ref{sec:skinae}) is that the network is not bound to a specific rig topology. We exploit this by introducing \textit{joint density} as an explicit conditional input. During training, we compute the normalized joint count of the ground truth asset. This scalar is embedded and injected into the flow model via AdaLN modulation~\citep{peebles2023scalable}.
At inference, users can adjust this scalar to control the granularity of the rig without changing the underlying geometry (visualized in Fig.~\ref{fig:joint-num}).

\begin{table*}
\centering
\caption{Quantitative evaluation of skeleton accuracy and skin metrics for various methods. Results include Chamfer distances, mm-space distances, Wasserstein-based distances, and skin metrics ($\ell_1$, $\ell_2$, KL divergence). The best scores are highlighted in bold, and the second-best scores are underlined. Note that TRELLIS$^*$ refers to the finetuned TRELLIS model trained on the split train set, while GT-Mesh combined with rigging methods serves as a reference.}
\label{tab:skeleton}
\resizebox{1.0\linewidth}{!}{
\begin{tabular}{l|ccc|cc|ccc}
\toprule
Method          & Joint-to-Joint $\downarrow$ & Joint-to-Bone $\downarrow$ & Bone-to-Bone $\downarrow$ & Wasserstein $\downarrow$ & Gromov–Wasserstein $\downarrow$ & Skin $\ell_1$ $\downarrow$ & Skin $\ell_2$ $\downarrow$ & Skin KL $\downarrow$ \\ \hline
\textcolor{gray}{GT-Mesh + UniRig}    & \textcolor{gray}{\underline{0.190}} & \textcolor{gray}{\underline{0.180}} & \textcolor{gray}{0.166} & \textcolor{gray}{0.260} & \textcolor{gray}{0.389} & \textcolor{gray}{0.0973} & \textcolor{gray}{0.224} & \textcolor{gray}{6.117}     \\ 
\textcolor{gray}{GT-Mesh + Anymate}    & \textcolor{gray}{\textbf{0.169}} & \textcolor{gray}{\textbf{0.161}} & \textcolor{gray}{\textbf{0.145}} & \textcolor{gray}{\textbf{0.232}} & \textcolor{gray}{\underline{0.368}} & \textcolor{gray}{\underline{0.0918}} & \textcolor{gray}{\underline{0.204}} & \textcolor{gray}{\underline{4.388}}     \\ 
\textcolor{gray}{GT-Mesh + Puppeteer}    & \textcolor{gray}{0.235} & \textcolor{gray}{0.228} & \textcolor{gray}{0.212} & \textcolor{gray}{\underline{0.239}} & \textcolor{gray}{\textbf{0.322}} & \textcolor{gray}{\textbf{0.0823}} & \textcolor{gray}{\textbf{0.197}} & \textcolor{gray}{\textbf{4.133}}      \\ 
\textcolor{gray}{GT-Mesh + RigAnything}   & \textcolor{gray}{0.193} & \textcolor{gray}{0.181} & \textcolor{gray}{\underline{0.165}}  & \textcolor{gray}{0.268}  & \textcolor{gray}{0.392} & \textcolor{gray}{0.1018} & \textcolor{gray}{0.223} & \textcolor{gray}{6.161}    \\ 
\hline
TRELLIS $^*$ + UniRig    & {0.205} & {0.192} & {0.179} & {0.269} & {0.397} & {0.0966} & {0.221} & {5.903}     \\ 
TRELLIS $^*$ + Anymate    & \underline{0.179} & \underline{0.172} & \underline{0.157} & \underline{0.232} & {0.349} & {0.0919} & {0.203} & {4.221}     \\ 
TRELLIS $^*$ + Puppeteer    & {0.245} & {0.237} & {0.219} & {0.241} & \underline{0.326} & \underline{0.0873} & \underline{0.202} & \underline{4.135}     \\
TRELLIS $^*$ + RigAnything   & {0.273} & {0.264} & {0.252}  & {0.285}  & {0.383} & {0.1026} & {0.224} & {6.451}   \\ 
\name & \cellcolor{yellow!50}\textbf{0.174} & \cellcolor{yellow!50}\textbf{0.164} & \cellcolor{yellow!50}\textbf{0.143} & \cellcolor{yellow!50}\textbf{0.229} & \cellcolor{yellow!50}\textbf{0.286} & \cellcolor{yellow!50}\textbf{0.0793} & \cellcolor{yellow!50}\textbf{0.186} & \cellcolor{yellow!50}\textbf{2.919} \\
\bottomrule
\end{tabular}
}
\end{table*}

\section{Experiment}

\subsection{Experimental Setup}

\subsubsection{Dataset and Implementation}

We adopt ArticulationXL~\cite{Song_2025_CVPR} as our evaluation dataset, which contains approximately 33K rigged 3D shapes curated from Objaverse-XL~\cite{deitke2023objaverse,deitke2023objaversexl}. We randomly sample 1K shapes to form the test set.
To increase motion diversity, we augment the dataset in two ways. For assets with existing animation sequences, we generate additional samples by interpolating within the asset’s own motion. Otherwise, we apply stochastic joint perturbations: each joint has 80\% probability of being rotated around a random axis by a jitter of up to $60^\circ$.
Because ArticulationXL is relatively small (in contrast to the $\sim$10M shapes in Objaverse-XL) and is insufficient for training a generative flow model from scratch, we adopt a warm-start initialization strategy. Specifically, we initialize the shared branches of the autoencoders and flow modules in \name with pre-trained TRELLIS parameters~\cite{xiang2025structured}, effectively leveraging large-scale geometric priors to facilitate the learning of joint representation.
\revise{In implementation, we first pre-train SkinAE and then freeze it when training the structured latent auto-encoder.}

\subsubsection{Baselines}

To the best of our knowledge, no existing method directly generates fully rigged 3D shapes -- comprising geometry, articulated skeleton, and skinning -- in a unified manner. Therefore, we construct strong baselines by coupling state-of-the-art automatic rigging methods with the recent 3D generative model TRELLIS~\cite{xiang2025structured}: we first generate a shape using TRELLIS, and then apply an off-the-shelf rigging algorithm to infer the skeleton and skinning weights. For a fair comparison under the same data domain and pose distribution, we further fine-tune TRELLIS on ArticulationXL using the same pose-augmented training set described above. We refer to this variant as TRELLIS$^*$ in Tab.\ref{tab:skeleton}. The rigging methods we evaluate in this pipeline include UniRig\cite{zhang2025one}, Anymate~\cite{deng2025anymate}, Puppeteer~\cite{songpuppeteer}, and RigAnything~\cite{liu2025riganything}.

\begin{figure*}
    \begin{center}
        \vspace{-0mm}
        \includegraphics[width=\linewidth]{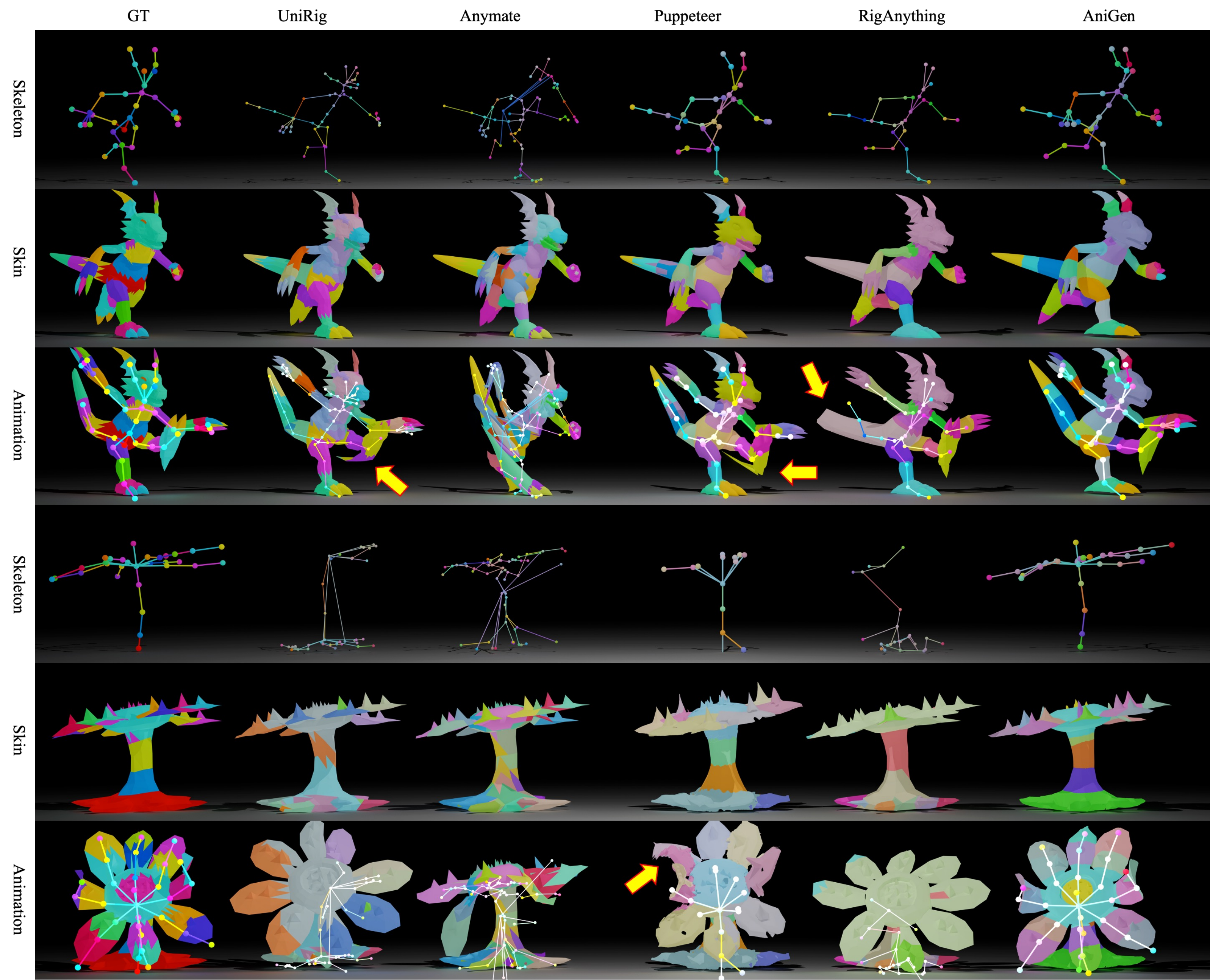}
    \end{center}
    \caption{Qualitative comparison of skeleton and skin results across different methods. We visualize the predicted skeletons, corresponding skins, and demonstrate animations to evaluate the practical usability of the rigged assets. The dragon case represents a relatively simple example with a clear identity and four-limbed structure, while the flower case poses a more complex challenge due to its intricate and non-standard structure.}
    \label{fig:comparison}
\end{figure*}

\begin{figure*}
    \begin{center}
        \vspace{-0mm}
        \includegraphics[width=\linewidth]{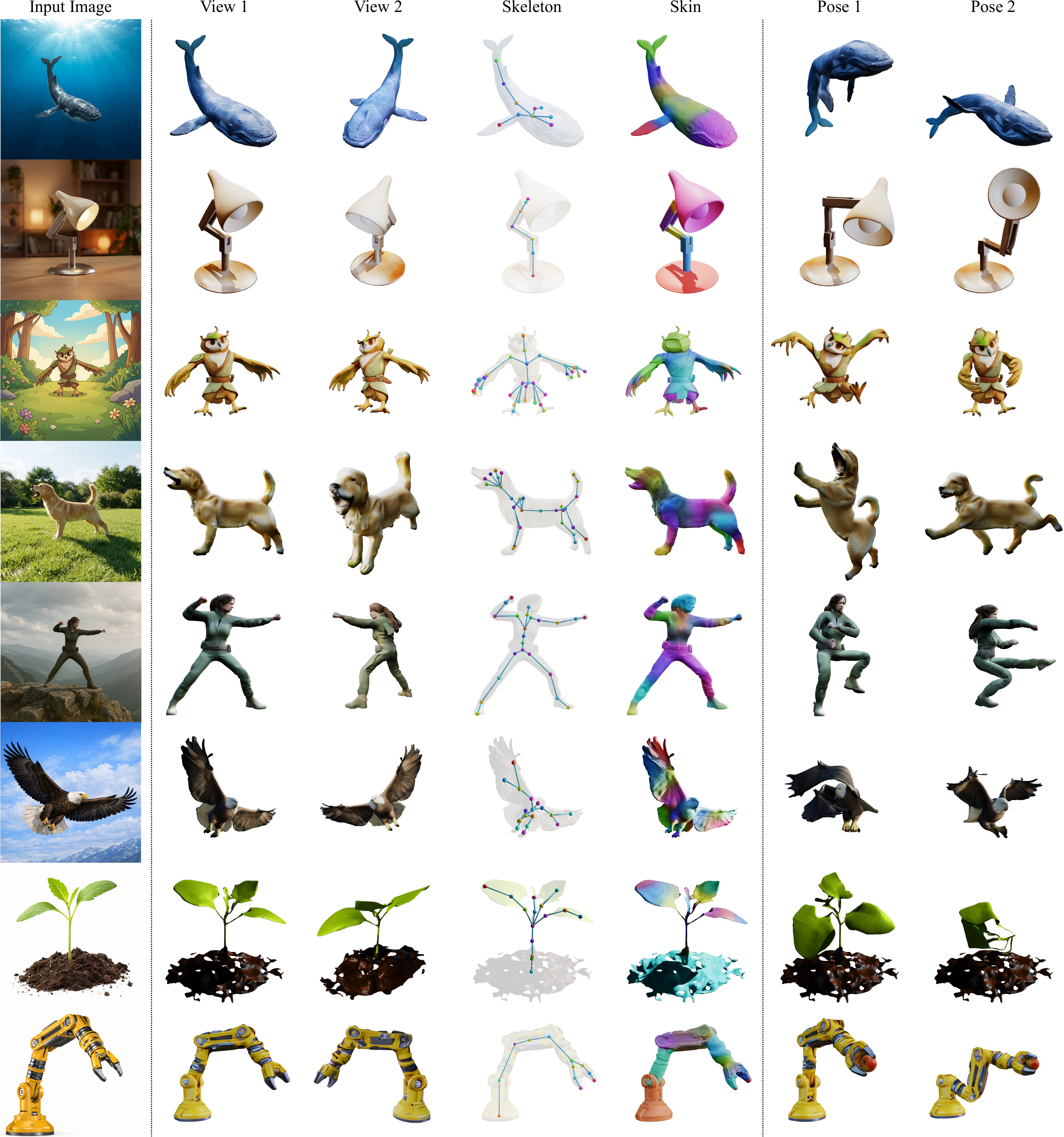}
    \end{center}
    \caption{Qualitative results on in-the-wild images, including examples from real-world photographs, web content, and AI-generated imagery. The showcased objects span a wide range of categories, such as sea animals, household items, cartoon characters, pets, humans, birds, plants, and machinery. Edited poses are demonstrated to illustrate animation capabilities, with examples such as a whale swimming, a lamp adjusting its angle, a dog running, and a robotic arm picking up an apple. These results highlight the versatility and practical usability of \name for applications in animation, games, image editing, VR, and more.}
    \label{fig:demo}
\end{figure*}

\subsubsection{Metrics}

Following UniRig~\cite{zhang2025one}, we adopt three Chamfer-style metrics to quantify skeletal geometric accuracy: \emph{joint-to-joint}, \emph{joint-to-bone}, and \emph{bone-to-bone} distances. While these metrics capture local Euclidean proximity, they are insufficient for comprehensive evaluation of skeleton structure.
In particular, they do not treat a skeleton as a \emph{metric measure space} and therefore can be insensitive to structural/topological errors. For example, inserting an extra joint along a ground-truth (GT) bone may not change the joint-to-bone or bone-to-bone distances, and adding a duplicated branch extremely close to a correct branch can still yield negligible Chamfer distances. This many-to-one matching bias inherently masks critical errors in kinematic connectivity and hierarchy.

To address this limitation, we incorporate Optimal Transport-based metrics, specifically the \emph{Wasserstein distance}~\cite{givens1984class} and \emph{Gromov--Wasserstein (GW) distance}~\cite{memoli2011gromov}, which explicitly account for the global distribution. Formally, we use the $L_2$-Wasserstein distance (Earth Mover's Distance) between joint measures:
\begin{equation}
\mathrm{D}_{W}(\mu,\nu)
=
\left(
\min_{\pi \in \Pi(a,b)}
\sum_{i=1}^{n}\sum_{k=1}^{m}
\pi_{ik}\,\|j_i - j^{gt}_k\|_2^{2}
\right)^{\frac{1}{2}},
\label{eq:wasserstein}
\end{equation}
where $\pi \in \mathbb{R}_{+}^{n \times m}$ is a transport plan with $\sum_{k}\pi_{ik}=\sum_{i}\pi_{ik}=1$.
Compared with Chamfer distances, the Wasserstein distance enforces a global mass-preserving matching, which mitigates many-to-one correspondences. However, it still relies on the ambient Euclidean cost and therefore does not explicitly encode skeletal topology.

To make the metric topology/structure-aware, we further compute the GW distance between the predicted skeleton graph and the GT skeleton graph by comparing \emph{intrinsic} pairwise distances. Let $d_{\mathrm{p}}(i,i')$ denote the geodesic distance along the predicted skeleton graph between predicted joints $i$ and $i'$, and let $d_{\mathrm{g}}(k,k')$ be defined analogously on the GT skeleton. The GW objective is
\begin{equation}
\mathrm{D}_{GW}(\mu,\nu)
=
\left(
\min_{\pi \in \Pi(a,b)}
\sum_{i,i'=1}^{n}\sum_{k,k'=1}^{m}
\big| d_{\mathrm{p}}(i,i') - d_{\mathrm{gt}}(k,k') \big|^{2}\,
\pi_{ik}\,\pi_{i'k'}
\right)^{\frac{1}{2}}.
\label{eq:gw}
\end{equation}
By matching geodesic structures rather than only Euclidean coordinates, GW penalizes topological inconsistencies such as spurious branches or incorrect connectivity even when the geometry is locally close.
In practice, the OT problems in Eq.~\eqref{eq:wasserstein} and Eq.~\eqref{eq:gw} can be efficiently solved via iterative Sinkhorn updates.
Given the optimal transport plan from Wasserstein distance, we align predicted skinning weights to GT and then report $\ell_1$, $\ell_2$, and KL divergence between the aligned skinning distributions.

\subsection{Quantitative Evaluation}

\revise{\paragraph{Rig Evaluation.}}
We summarize the quantitative evaluation results in Tab.~\ref{tab:skeleton}, including Chamfer distances, mm-space distances, and skin metrics. Note that TRELLIS$^*$ refers to the finetuned TRELLIS model trained on the split train set.
\revise{Because every method is image-conditioned rather than GT-conditioned, the generated asset is not expected to be an exact replica of the GT shape. We therefore normalize scale, center the prediction, and then align it to the GT with ICP using 100 randomly initialized rotations.}
The results demonstrate that \name achieves the best performance in skeleton structure prediction and skin accuracy across all baselines. This establishes \name as the leading end-to-end image-conditioned rigged shape generation solution. Notably, \name excels in Gromov-Wasserstein distance and skin KL divergence, achieving significant advantages over the other baselines in the accuracy of skeleton topology and skin weights. Additionally, we provide results from coupling GT meshes with off-the-shelf rigging methods to serve as an upper-bound reference. While these GT-input baselines naturally exhibit better results, it is worth noting that generation models often produce geometries that deviate slightly from the GT due to minor variations in scale, rotation, and pose.

\revise{\paragraph{Geometry Evaluation.}
We further report geometry-and-fidelity metrics in Tab.~\ref{tab:geometry}. \name remains competitive with TRELLIS$^*$, it shows only a small gap, while substantially improving the rigging-related metrics in Tab.~\ref{tab:skeleton}. This confirms that jointly modeling shape, skeleton, and skin does not substantially degrade geometry quality. Pure geometry generation can still be slightly stronger under geometry-only metrics, which we consider a minor limitation, but the gap is small relative to the articulation gains.}

\begin{table}[h]
\centering
\caption{\revise{Geometry evaluation on the rigged-domain test set. We report surface Chamfer distance, F-score, and image-space PSNR \& LPIPS.}}
\label{tab:geometry}
\resizebox{0.75\linewidth}{!}{
\begin{tabular}{l|cccc}
\toprule
\revise{Method} & \revise{Chamfer $\downarrow$} & \revise{F-Score $\uparrow$} & \revise{PSNR $\uparrow$} & \revise{LPIPS $\downarrow$} \\
\hline
\revise{TRELLIS} & \revise{0.0475} & \revise{0.77} & \revise{24.25} & \revise{0.126} \\
\revise{TRELLIS$^*$} & \revise{\textbf{0.0394}} & \revise{\textbf{0.89}} & \revise{\textbf{25.23}} & \revise{\textbf{0.104}} \\
\revise{\name} & \revise{\underline{0.0409}} & \revise{\underline{0.88}} & \revise{\underline{25.12}} & \revise{\underline{0.108}} \\
\bottomrule
\end{tabular}
}
\end{table}

\revise{\paragraph{Inference Cost Evaluation.}
We report end-to-end inference cost in Tab.~\ref{tab:runtime}. \name is comparable to the fastest sequential baseline while avoiding the heavy post-hoc rigging overhead of methods such as UniRig and RigAnything.}

\begin{table}[h]
\centering
\caption{\revise{Inference cost runtime comparison.}}
\label{tab:runtime}
\resizebox{1.0\linewidth}{!}{
\begin{tabular}{l|ccc}
\toprule
\revise{Method} & \revise{TRELLIS$^*$} & \revise{TRELLIS$^*$+UniRig} & \revise{TRELLIS$^*$+Anymate} \\
\hline
\revise{Time (s) $\downarrow$} & \revise{15} & \revise{146} & \revise{19} \\
\hline
\revise{Method} & \revise{TRELLIS$^*$+Puppeteer} & \revise{TRELLIS$^*$+RigAnything} & \revise{\name} \\
\hline
\revise{Time (s) $\downarrow$} & \revise{36} & \revise{127} & \revise{19} \\
\bottomrule
\end{tabular}
}
\end{table}

\subsection{Qualitative Evaluation}

To provide a more intuitive comparison, we present qualitative results across various baselines in Fig.~\ref{fig:comparison}. We visualize the generated skeletons and skins and perform similar animations on the outputs to better demonstrate the practical usability of each rigged asset. For brevity, we omit TRELLIS$^*$+ in the following discussion.

In the case of the dragon, which is a relatively simple example with a clear identity and a four-limbed structure, \name, UniRig, Puppeteer, and RigAnything generate skeletons that are overall very similar to the ground truth. However, there are notable differences in the details. UniRig generates redundant bones in the head, while Puppeteer and RigAnything fail to produce detailed finger joints. Anymate produces a skeleton with joint distributions very close to the GT, but the connections between bones are incorrect. Regarding skin results, UniRig, Puppeteer, and RigAnything exhibit regional artifacts, particularly in the feet or tail.

The flower example is more challenging than the dragon. UniRig and RigAnything fail to generate skeletons that adequately cover the full structure of the flower. In contrast, Anymate produces joints that closely match the GT joint distribution, but some bone connections are incorrect. Puppeteer creates a coarse yet overall suitable skeleton to support the flower, but its skin results are inadequate for practical animation, resulting in broken animations.
\revise{Some compared results also exhibit pose discrepancies after animation. This is not because different target poses are used; instead, all methods are driven toward the same target motion. When a baseline predicts topologically broken bones or erroneous skin influences, it cannot physically realize the target pose without catastrophic distortion, so the final animated pose remains visibly mismatched.}

\subsection{In-the-Wild Results}

We present the robust generalization of \name on diverse ``in-the-wild'' images in Fig.~\ref{fig:demo}, including real-world photographs, web-sourced imagery, and AI-generated content. These examples span a diverse set of object categories: sea animals, household items, cartoon characters, pets, humans, birds, plants, and machinery, demonstrating the versatility of our approach across both natural and synthetic visual domains.
To highlight the functional utility of the generated assets, we further provide edited poses and motion variants tailored to the identity, structure, and expected behavior of the underlying subject. As illustrated, a whale can be posed to swim freely through an ocean scene; the lamp can be reoriented so its head and body direct light toward different targets; and the cartoon character can be driven through a variety of full-body motions. Likewise, the dog can open and close its mouth while running across a lawn, the woman can be animated performing kung-fu movements, and the eagle can be posed to hug or capture a sheep in a dynamic interaction. Beyond animals and humans, we also demonstrate controllable state changes and functional motions: the plant can transition between blooming and withering, and the robotic arm can grasp an apple, lift it, and transport it to a new location.

These diverse results underscore the flexibility and versatility of \name, demonstrating its capacity to operate effectively and robustly across a wide variety of subjects, visual styles, and real-world scenarios. This breadth suggests that \name serves as a unified, category-agnostic foundation for controllable asset synthesis and animation, rather than being confined to narrow context domains.
As a result, \name enables a wide range of downstream applications, including embodied AI (e.g., interactive agents that require consistent, controllable visual assets), image and video editing (e.g., pose- and motion-aware content modification), animation and gaming pipelines (e.g., rapid prototyping of characters, props, and actions), and creative production workflows such as cartoon creation and stylized storytelling. Moreover, it can support immersive and simulation-centric settings, including virtual reality experiences, digital-twin systems, and game character development.

\begin{figure}[h]
    \begin{center}
        \vspace{-0mm}
        \includegraphics[width=1.0\linewidth]{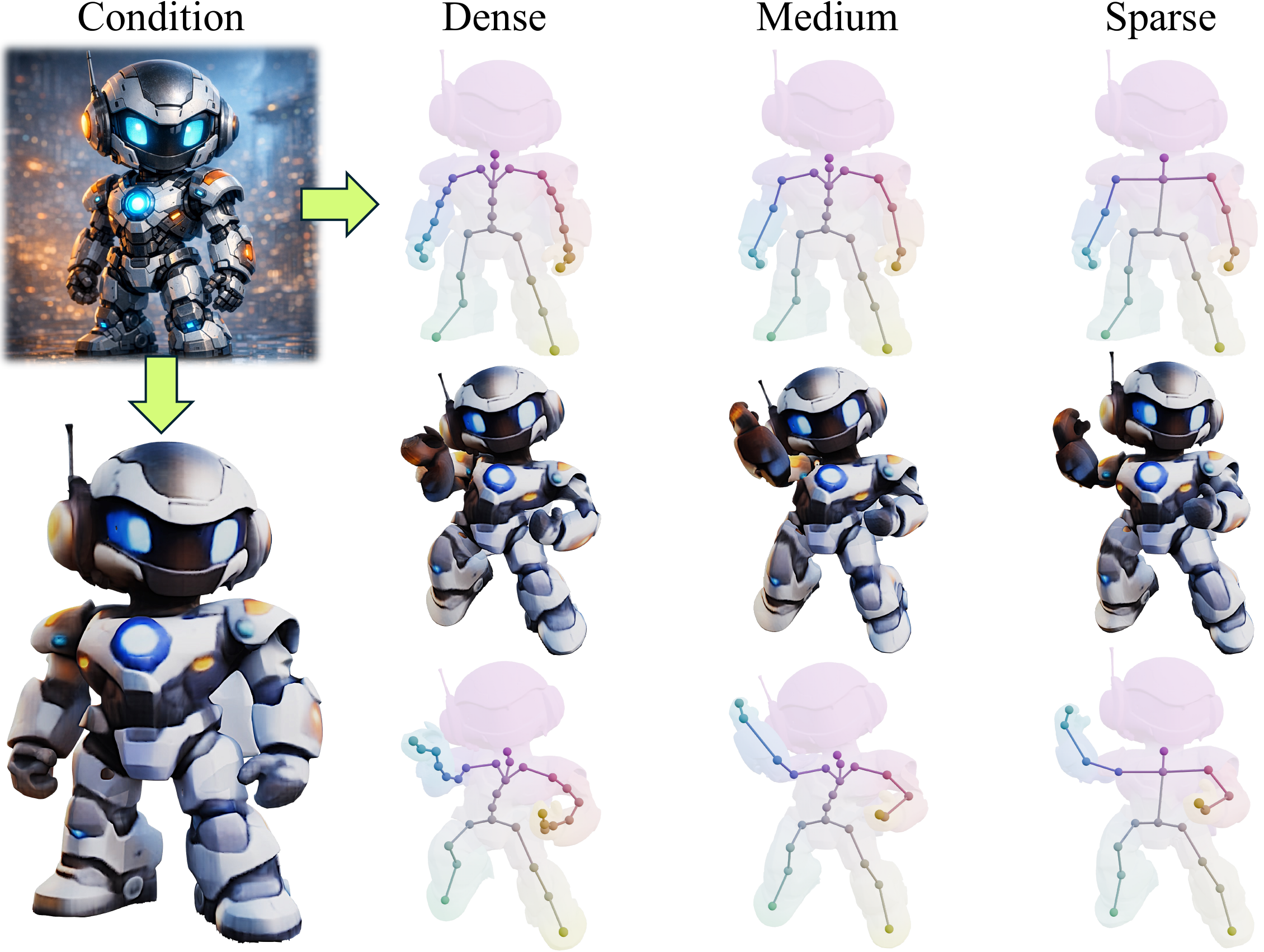}
    \end{center}
    \caption{Skeleton generation with different joint density levels. Higher joint density enables flexible, human-like motions, while medium and sparse densities result in reduced flexibility, resembling real robots. This demonstrates the adaptability of the method to varying motion requirements.}
    \label{fig:joint-num}
\end{figure}

\subsection{Joint Number Control}

As discussed in Sec.~\ref{sec:slat-flow}, we introduce joint density as a conditional input to control the number of joints in the generated skeleton, allowing it to adapt to varying flexibility requirements. The GT joint count is normalized to the range $[0,1]$ (by dividing by 60 and clamping), positionally embedded, and encoded with MLPs before being injected into the flow model via AdaLN modulation. During inference, the joint density condition can be adjusted to control the final number of skeleton joints using classifier-free guidance (CFG).
We illustrate the results of joint number control in Fig.~\ref{fig:joint-num}, using a CFG scale of 3.0. By adjusting the joint density, the model can synthesize skeletons with distinct degrees of freedom: a ``high-density'' can perform smooth, human-like motions -- such as bending arms, twisting its head, and clenching its fists -- while a ``medium-density'' variant retains limb flexibility but loses fine-grained manual dexterity. At the ``sparse'' extreme, the model yields a rigid, minimalist armature with significantly constrained motion, like a ``real robot''.

\subsection{Ablation Study}
\label{sec:ablation}

\begin{figure}[h]
    \begin{center}
        \vspace{-0mm}
        \includegraphics[width=1.0\linewidth]{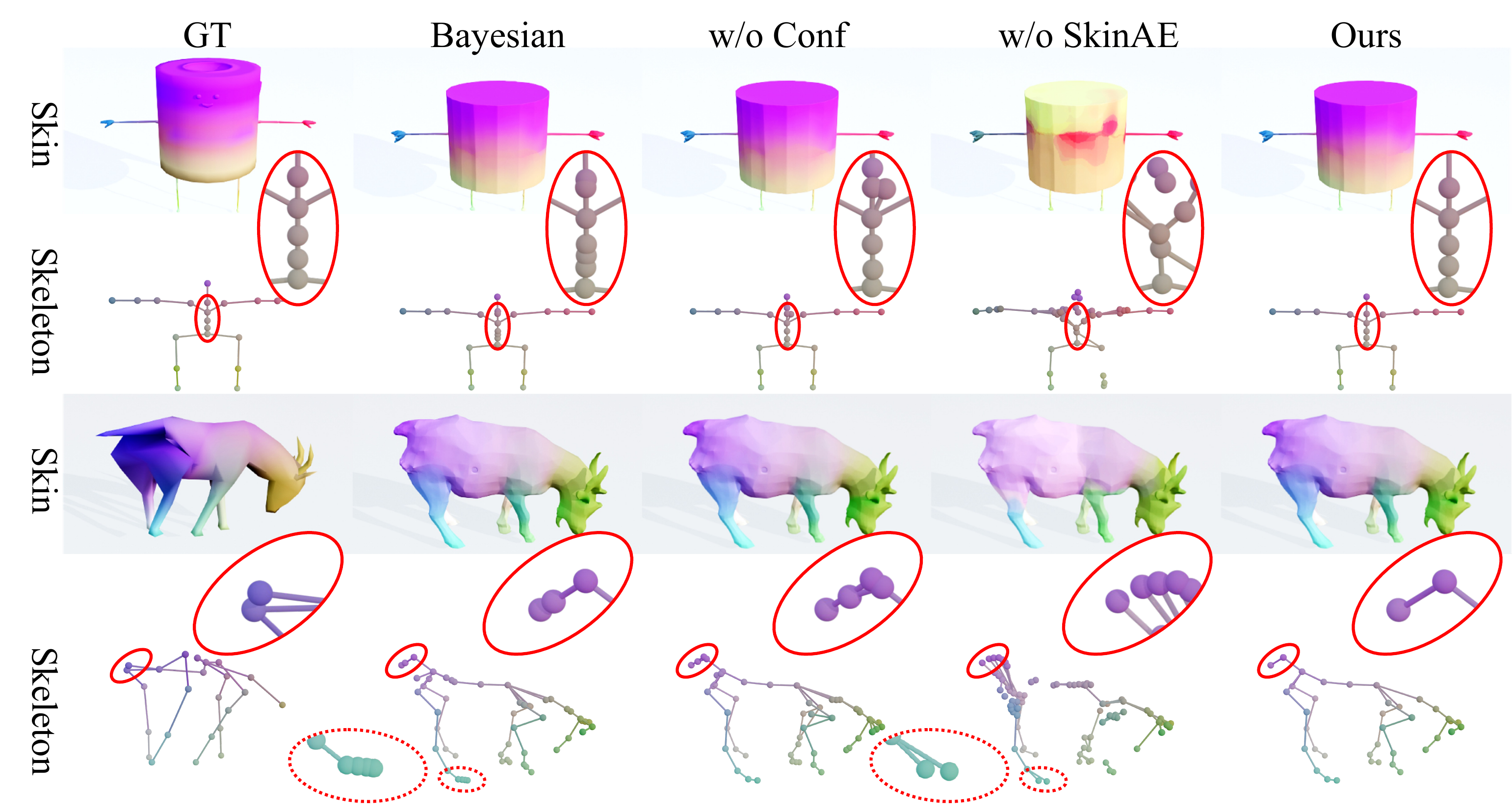}
    \end{center}
    \caption{Ablation study on confidence design and SkinAE pretraining. Without confidence learning or with Bayesian confidence, skeleton predictions are noisy and redundant (column 2,3). Without pretraining SkinAE, skin quality and skeleton structure degrade significantly (column 4). Our method ensures clean skeletons and accurate skin generation (column 5).}
    \label{fig:ablation_conf_exp}
\end{figure}

In this section, we conduct an ablation study on the design choices discussed in the method section and explain why we selected the current technical approach.
First, we investigate the confidence design, as detailed in Sec.\ref{sec:skeleton-field}. High ambiguity in border regions makes confidence learning essential to ensure clean final generation results. Without confidence, the model fails to refine noisy predictions into a clean skeleton, as shown in the 3rd column of Fig.\ref{fig:ablation_conf_exp}. Even with Bayesian confidence~\cite{kendall2017uncertainties}, the self-adaptive learning approach cannot effectively resolve ambiguous regions, resulting in noisy and redundant bones and joints.
In contrast, our method explicitly defines ambiguity in Eq.\ref{eq:confidence} using a prior and supervises the confidence field directly during training, rather than relying on Bayesian learning's adaptive reconstruction loss. This results in a more effective confidence field, which integrates seamlessly with the confidence-weighted grouping algorithm (Alg.\ref{alg:meanshift_grouping}) to consistently achieve accurate grouping results.
\revise{The quantitative results in Tab.~\ref{tab:ablation} confirm this trend: explicit confidence supervision improves structural correctness over both removing confidence and replacing it with Bayesian uncertainty learning, while SkinAE pretraining is critical for high-quality skin prediction.}

\begin{table}[h]
\centering
\caption{\revise{Quantitative ablation on confidence modeling and SkinAE pretraining. Lower is better for both metrics.}}
\label{tab:ablation}
\resizebox{0.85\linewidth}{!}{
\begin{tabular}{l|cccc}
\toprule
\revise{Method} & \revise{Bayesian} & \revise{w/o-Conf} & \revise{w/o-SkinAE} & \revise{Ours} \\
\hline
\revise{Joint-GW $\downarrow$} & \revise{0.310} & \revise{0.337} & \revise{0.383} & \revise{\textbf{0.286}} \\
\revise{Skin-KL $\downarrow$} & \revise{3.174} & \revise{3.187} & \revise{5.138} & \revise{\textbf{2.919}} \\
\bottomrule
\end{tabular}
}
\end{table}

We also analyze the impact of pretraining SkinAE. Without pretraining ("w/o SkinAE"), SkinAE becomes part of the structured latent auto-encoder ($\mathcal{E}_L$ and $\mathcal{D}_L$) and is trained jointly from scratch. This joint optimization lacks skin information in the input to the structured latent encoder, leading to sub-optimal feature alignment and hindered convergence. 
Consequently, the reconstruction of the joint field is negatively affected. As shown in the 4th column of Fig.~\ref{fig:ablation_conf_exp}, this results in poor skin quality for the tissue character and a broken skeleton for the goat. Our pretraining strategy ensures convergence of the structured latent auto-encoder and achieves sufficient accuracy for the generation decoding task.

\section{Conclusion}
We introduced \name, a method for image-conditioned generation of animatable 3D assets that couples geometry synthesis with rig prediction in a single generative model. In contrast to ``generate-then-rig'' pipelines, \name learns a joint distribution over shape, skeleton, and skin, and represents all three as fields to promote mutual consistency. Our core contribution, the $S^3$ field formulation, together with the confidence-decaying nearest joint-parent field and the dual skin feature field, enables reliable modeling of kinematic structure and skinning despite ambiguity and category diversity. \name leverages a structured latent auto-encoder and flow-based generation over sparse structures and structured latents to produce coherent rigged assets directly from images.
Empirically, \name achieves clear gains over prior approaches in rigging quality and usability, and generalizes well to in-the-wild images spanning a wide range of object categories and visual styles. \revise{The experiments further show that these gains are achieved without sacrificing geometry fidelity relative to strong geometry-only generators.} We believe that \name establishes a powerful foundation for the next generation of controllable 3D content creation, with broad applications for interactive graphics, embodied AI, VR/AR, digital twins, and animation/editing workflows.

\revise{
\paragraph{Limitations and future work.}
One limitation of \name is its current focus on image-conditioned generation. Although effective, this setting does not fully reflect many practical use cases, where animatable shapes are often derived from captured videos in which motion dynamics and skeletal constraints are more explicitly observed. Extending \name to support video input could enable more robust and temporally consistent shape and skeleton generation, and also enable the production of animatable shapes directly aligned with the motions and structural cues present in the input video. Addressing this challenge is an important direction for future work and could substantially broaden the applicability of the method.
A second limitation arises for articulated objects that require strict geometric alignment between rigid parts, such as the lid and base of a laptop. Although \name can infer a plausible hinge skeleton for such objects, small geometric mismatches may still produce visible gaps in closed configurations.
Finally, the skeletons predicted by our current model primarily reflect the medial-axis tendencies present in the training data, rather than fully anatomically inspired production rigs. We emphasize that this limitation stems from the available data rather than from the representation itself. Since the proposed $S^3$ fields are continuous and volumetric, they are, in principle, capable of encoding richer sub-surface control structures, including production-style anatomical rigs, given access to suitable data.
}

{
\small
\bibliographystyle{ACM-Reference-Format}
\bibliography{main}

@inproceedings{jain2022zero,
  title={Zero-shot text-guided object generation with dream fields},
  author={Jain, Ajay and Mildenhall, Ben and Barron, Jonathan T and Abbeel, Pieter and Poole, Ben},
  booktitle={Proceedings of the IEEE/CVF conference on computer vision and pattern recognition},
  pages={867--876},
  year={2022}
}

@inproceedings{poole2023dreamfusion,
  title={DreamFusion: Text-to-3D using 2D Diffusion},
  author={Poole, Ben and Jain, Ajay and Barron, Jonathan T and Mildenhall, Ben},
  booktitle={ICLR},
  year={2023}
}

@article{TripoSR2024,
  title={TripoSR: Fast 3D Object Reconstruction from a Single Image},
  author={Tochilkin, Dmitry and Pankratz, David and Liu, Zexiang and Huang, Zixuan and and Letts, Adam and Li, Yangguang and Liang, Ding and Laforte, Christian and Jampani, Varun and Cao, Yan-Pei},
  journal={arXiv preprint arXiv:2403.02151},
  year={2024}
}

@inproceedings{zou2024triplane,
  title={Triplane meets gaussian splatting: Fast and generalizable single-view 3d reconstruction with transformers},
  author={Zou, Zi-Xin and Yu, Zhipeng and Guo, Yuan-Chen and Li, Yangguang and Liang, Ding and Cao, Yan-Pei and Zhang, Song-Hai},
  booktitle={Proceedings of the IEEE/CVF conference on computer vision and pattern recognition},
  pages={10324--10335},
  year={2024}
}

@inproceedings{xiang2025structured,
  title={Structured 3d latents for scalable and versatile 3d generation},
  author={Xiang, Jianfeng and Lv, Zelong and Xu, Sicheng and Deng, Yu and Wang, Ruicheng and Zhang, Bowen and Chen, Dong and Tong, Xin and Yang, Jiaolong},
  booktitle={Proceedings of the Computer Vision and Pattern Recognition Conference},
  pages={21469--21480},
  year={2025}
}

@article{xiang2025native,
  title={Native and Compact Structured Latents for 3D Generation},
  author={Xiang, Jianfeng and Chen, Xiaoxue and Xu, Sicheng and Wang, Ruicheng and Lv, Zelong and Deng, Yu and Zhu, Hongyuan and Dong, Yue and Zhao, Hao and Yuan, Nicholas Jing and others},
  journal={arXiv preprint arXiv:2512.14692},
  year={2025}
}

@InProceedings{He_2025_ICCV,
    author    = {He, Xianglong and Zou, Zi-Xin and Chen, Chia-Hao and Guo, Yuan-Chen and Liang, Ding and Yuan, Chun and Ouyang, Wanli and Cao, Yan-Pei and Li, Yangguang},
    title     = {SparseFlex: High-Resolution and Arbitrary-Topology 3D Shape Modeling},
    booktitle = {Proceedings of the IEEE/CVF International Conference on Computer Vision (ICCV)},
    month     = {October},
    year      = {2025},
    pages     = {14822-14833}
}

@article{chen2025ultra3d,
  title={Ultra3d: Efficient and high-fidelity 3d generation with part attention},
  author={Chen, Yiwen and Li, Zhihao and Wang, Yikai and Zhang, Hu and Li, Qin and Zhang, Chi and Lin, Guosheng},
  journal={arXiv preprint arXiv:2507.17745},
  year={2025}
}

@article{wu2024direct3d,
  title={Direct3d: Scalable image-to-3d generation via 3d latent diffusion transformer},
  author={Wu, Shuang and Lin, Youtian and Zhang, Feihu and Zeng, Yifei and Xu, Jingxi and Torr, Philip and Cao, Xun and Yao, Yao},
  journal={Advances in Neural Information Processing Systems},
  volume={37},
  pages={121859--121881},
  year={2024}
}

@article{zhang20233dshape2vecset,
  title={3dshape2vecset: A 3d shape representation for neural fields and generative diffusion models},
  author={Zhang, Biao and Tang, Jiapeng and Niessner, Matthias and Wonka, Peter},
  journal={ACM Transactions On Graphics (TOG)},
  volume={42},
  number={4},
  pages={1--16},
  year={2023},
  publisher={ACM New York, NY, USA}
}

@inproceedings{chen2025dora,
  title={Dora: Sampling and benchmarking for 3d shape variational auto-encoders},
  author={Chen, Rui and Zhang, Jianfeng and Liang, Yixun and Luo, Guan and Li, Weiyu and Liu, Jiarui and Li, Xiu and Long, Xiaoxiao and Feng, Jiashi and Tan, Ping},
  booktitle={Proceedings of the Computer Vision and Pattern Recognition Conference},
  pages={16251--16261},
  year={2025}
}

@article{zhang2024clay,
  title={Clay: A controllable large-scale generative model for creating high-quality 3d assets},
  author={Zhang, Longwen and Wang, Ziyu and Zhang, Qixuan and Qiu, Qiwei and Pang, Anqi and Jiang, Haoran and Yang, Wei and Xu, Lan and Yu, Jingyi},
  journal={ACM Transactions on Graphics (TOG)},
  volume={43},
  number={4},
  pages={1--20},
  year={2024},
  publisher={ACM New York, NY, USA}
}

@article{li2025triposg,
  title={Triposg: High-fidelity 3d shape synthesis using large-scale rectified flow models},
  author={Li, Yangguang and Zou, Zi-Xin and Liu, Zexiang and Wang, Dehu and Liang, Yuan and Yu, Zhipeng and Liu, Xingchao and Guo, Yuan-Chen and Liang, Ding and Ouyang, Wanli and others},
  journal={arXiv preprint arXiv:2502.06608},
  year={2025}
}

@inproceedings{honglrm,
  title={LRM: Large Reconstruction Model for Single Image to 3D},
  author={Hong, Yicong and Zhang, Kai and Gu, Jiuxiang and Bi, Sai and Zhou, Yang and Liu, Difan and Liu, Feng and Sunkavalli, Kalyan and Bui, Trung and Tan, Hao},
  booktitle={The Twelfth International Conference on Learning Representations},
  year={2023}
}

@article{kendall2017uncertainties,
  title={What uncertainties do we need in bayesian deep learning for computer vision?},
  author={Kendall, Alex and Gal, Yarin},
  journal={Advances in neural information processing systems},
  volume={30},
  year={2017}
}

@article{shen2023flexible,
  title={Flexible isosurface extraction for gradient-based mesh optimization},
  author={Shen, Tianchang and Munkberg, Jacob and Hasselgren, Jon and Yin, Kangxue and Wang, Zian and Chen, Wenzheng and Gojcic, Zan and Fidler, Sanja and Sharp, Nicholas and Gao, Jun},
  journal={ACM Transactions on Graphics (TOG)},
  volume={42},
  number={4},
  pages={1--16},
  year={2023},
  publisher={ACM New York, NY, USA}
}

@InProceedings{Wu_2025_ICCV,
    author    = {Wu, Zijie and Yu, Chaohui and Wang, Fan and Bai, Xiang},
    title     = {AnimateAnyMesh: A Feed-Forward 4D Foundation Model for Text-Driven Universal Mesh Animation},
    booktitle = {Proceedings of the IEEE/CVF International Conference on Computer Vision (ICCV)},
    month     = {October},
    year      = {2025},
    pages     = {13557-13568}
}

@article{xu2020rignet,
  title={RigNet: neural rigging for articulated characters},
  author={Xu, Zhan and Zhou, Yang and Kalogerakis, Evangelos and Landreth, Chris and Singh, Karan},
  journal={ACM Transactions on Graphics (TOG)},
  volume={39},
  number={4},
  pages={58--1},
  year={2020},
  publisher={ACM New York, NY, USA}
}

@article{zhang2025one,
  title={One model to rig them all: Diverse skeleton rigging with unirig},
  author={Zhang, Jia-Peng and Pu, Cheng-Feng and Guo, Meng-Hao and Cao, Yan-Pei and Hu, Shi-Min},
  journal={ACM Transactions on Graphics (TOG)},
  volume={44},
  number={4},
  pages={1--18},
  year={2025},
  publisher={ACM New York, NY, USA}
}

@inproceedings{deng2025anymate,
  title={Anymate: A dataset and baselines for learning 3d object rigging},
  author={Deng, Yufan and Zhang, Yuhao and Geng, Chen and Wu, Shangzhe and Wu, Jiajun},
  booktitle={Proceedings of the Special Interest Group on Computer Graphics and Interactive Techniques Conference Conference Papers},
  pages={1--10},
  year={2025}
}

@article{liu2025riganything,
  title={Riganything: Template-free autoregressive rigging for diverse 3d assets},
  author={Liu, Isabella and Xu, Zhan and Yifan, Wang and Tan, Hao and Xu, Zexiang and Wang, Xiaolong and Su, Hao and Shi, Zifan},
  journal={ACM Transactions on Graphics (TOG)},
  volume={44},
  number={4},
  pages={1--12},
  year={2025},
  publisher={ACM New York, NY, USA}
}

@inproceedings{songpuppeteer,
  title={Puppeteer: Rig and Animate Your 3D Models},
  author={Song, Chaoyue and Li, Xiu and Yang, Fan and Xu, Zhongcong and Wei, Jiacheng and Liu, Fayao and Feng, Jiashi and Lin, Guosheng and Zhang, Jianfeng},
  booktitle={The Thirty-ninth Annual Conference on Neural Information Processing Systems},
  year={2025},
}

@article{ren2023dreamgaussian4d,
  title={Dreamgaussian4d: Generative 4d gaussian splatting},
  author={Ren, Jiawei and Pan, Liang and Tang, Jiaxiang and Zhang, Chi and Cao, Ang and Zeng, Gang and Liu, Ziwei},
  journal={arXiv preprint arXiv:2312.17142},
  year={2023}
}

@inproceedings{wu2024sc4d,
  title={Sc4d: Sparse-controlled video-to-4d generation and motion transfer},
  author={Wu, Zijie and Yu, Chaohui and Jiang, Yanqin and Cao, Chenjie and Wang, Fan and Bai, Xiang},
  booktitle={European Conference on Computer Vision},
  pages={361--379},
  year={2024},
  organization={Springer}
}

@inproceedings{kim2025rigidity,
  title={Rigidity-Aware 3D Gaussian Deformation from a Single Image},
  author={Kim, Jinhyeok and Bang, Jaehun and Seo, Seunghyun and Joo, Kyungdon},
  booktitle={Proceedings of the SIGGRAPH Asia 2025 Conference Papers},
  pages={1--11},
  year={2025}
}

@inproceedings{huang2024sc,
  title={Sc-gs: Sparse-controlled gaussian splatting for editable dynamic scenes},
  author={Huang, Yi-Hua and Sun, Yang-Tian and Yang, Ziyi and Lyu, Xiaoyang and Cao, Yan-Pei and Qi, Xiaojuan},
  booktitle={Proceedings of the IEEE/CVF conference on computer vision and pattern recognition},
  pages={4220--4230},
  year={2024}
}

@inproceedings{wu2025cat4d,
  title={Cat4d: Create anything in 4d with multi-view video diffusion models},
  author={Wu, Rundi and Gao, Ruiqi and Poole, Ben and Trevithick, Alex and Zheng, Changxi and Barron, Jonathan T and Holynski, Aleksander},
  booktitle={Proceedings of the Computer Vision and Pattern Recognition Conference},
  pages={26057--26068},
  year={2025}
}

@article{xie2025animamimic,
  title={AnimaMimic: Imitating 3D Animation from Video Priors},
  author={Xie, Tianyi and Chen, Yunuo and Guo, Yaowei and Yang, Yin and Zhou, Bolei and Terzopoulos, Demetri and Jiang, Ying and Jiang, Chenfanfu},
  journal={arXiv preprint arXiv:2512.14133},
  year={2025}
}

@article{baran2007automatic,
  title={Automatic rigging and animation of 3d characters},
  author={Baran, Ilya and Popovi{\'c}, Jovan},
  journal={ACM Transactions on graphics (TOG)},
  volume={26},
  number={3},
  pages={72--es},
  year={2007},
  publisher={ACM New York, NY, USA}
}

@article{marr1978representation,
  title={Representation and recognition of the spatial organization of three-dimensional shapes},
  author={Marr, David and Nishihara, Herbert Keith},
  journal={Proceedings of the Royal Society of London. Series B. Biological Sciences},
  volume={200},
  number={1140},
  pages={269--294},
  year={1978},
  publisher={The Royal Society London}
}

@article{zhang2024skinned,
  title={Skinned motion retargeting with preservation of body part relationships},
  author={Zhang, Jia-Qi and Wang, Miao and Zhang, Fu-Cheng and Zhang, Fang-Lue},
  journal={IEEE Transactions on Visualization and Computer Graphics},
  year={2024},
  publisher={IEEE}
}

@article{wang2023hmc,
  title={Hmc: Hierarchical mesh coarsening for skeleton-free motion retargeting},
  author={Wang, Haoyu and Huang, Shaoli and Zhao, Fang and Yuan, Chun and Shan, Ying},
  journal={arXiv preprint arXiv:2303.10941},
  year={2023}
}

@inproceedings{xu2022morig,
  title={Morig: Motion-aware rigging of character meshes from point clouds},
  author={Xu, Zhan and Zhou, Yang and Yi, Li and Kalogerakis, Evangelos},
  booktitle={SIGGRAPH Asia 2022 conference papers},
  pages={1--9},
  year={2022}
}

@article{li2021learning,
  author = {Li, Peizhuo and Aberman, Kfir and Hanocka, Rana and Liu, Libin and Sorkine-Hornung, Olga and Chen, Baoquan},
  title = {Learning Skeletal Articulations with Neural Blend Shapes},
  journal = {ACM Transactions on Graphics (TOG)},
  volume = {40},
  number = {4},
  pages = {130},
  year = {2021},
  publisher = {ACM}
}

@article{liu2019neuroskinning,
  title={Neuroskinning: Automatic skin binding for production characters with deep graph networks},
  author={Liu, Lijuan and Zheng, Youyi and Tang, Di and Yuan, Yi and Fan, Changjie and Zhou, Kun},
  journal={ACM Transactions on Graphics (ToG)},
  volume={38},
  number={4},
  pages={1--12},
  year={2019},
  publisher={ACM New York, NY, USA}
}

@article{ma2023tarig,
  title={TARig: Adaptive template-aware neural rigging for humanoid characters},
  author={Ma, Jing and Zhang, Dongliang},
  journal={Computers \& Graphics},
  volume={114},
  pages={158--167},
  year={2023},
  publisher={Elsevier}
}

@inproceedings{sun2025drive,
  title={Drive: Diffusion-based rigging empowers generation of versatile and expressive characters},
  author={Sun, Mingze and Chen, Junhao and Dong, Junting and Chen, Yurun and Jiang, Xinyu and Mao, Shiwei and Jiang, Puhua and Wang, Jingbo and Dai, Bo and Huang, Ruqi},
  booktitle={Proceedings of the Computer Vision and Pattern Recognition Conference},
  pages={21170--21180},
  year={2025}
}

@inproceedings{tang2024lgm,
  title={Lgm: Large multi-view gaussian model for high-resolution 3d content creation},
  author={Tang, Jiaxiang and Chen, Zhaoxi and Chen, Xiaokang and Wang, Tengfei and Zeng, Gang and Liu, Ziwei},
  booktitle={European Conference on Computer Vision},
  pages={1--18},
  year={2024},
  organization={Springer}
}

@article{kerbl20233d,
  title={3D Gaussian splatting for real-time radiance field rendering.},
  author={Kerbl, Bernhard and Kopanas, Georgios and Leimk{\"u}hler, Thomas and Drettakis, George},
  journal={ACM Trans. Graph.},
  volume={42},
  number={4},
  pages={139--1},
  year={2023}
}

@inproceedings{gat2025anytop,
  title={Anytop: Character animation diffusion with any topology},
  author={Gat, Inbar and Raab, Sigal and Tevet, Guy and Reshef, Yuval and Bermano, Amit Haim and Cohen-Or, Daniel},
  booktitle={Proceedings of the Special Interest Group on Computer Graphics and Interactive Techniques Conference Conference Papers},
  pages={1--10},
  year={2025}
}

@inproceedings{huang2025animax,
  title={Animax: Animating the inanimate in 3d with joint video-pose diffusion models},
  author={Huang, Zehuan and Feng, Haoran and Sun, Yang-Tian and Guo, Yuan-Chen and Cao, Yan-Pei and Sheng, Lu},
  booktitle={Proceedings of the SIGGRAPH Asia 2025 Conference Papers},
  pages={1--13},
  year={2025}
}

@article{guo2025make,
  title={Make-It-Poseable: Feed-forward Latent Posing Model for 3D Humanoid Character Animation},
  author={Guo, Zhiyang and Zhang, Ori and Xiang, Jax and Zhao, Alan and Zhou, Wengang and Li, Houqiang},
  journal={arXiv preprint arXiv:2512.16767},
  year={2025}
}

@inproceedings{radford2021learning,
  title={Learning transferable visual models from natural language supervision},
  author={Radford, Alec and Kim, Jong Wook and Hallacy, Chris and Ramesh, Aditya and Goh, Gabriel and Agarwal, Sandhini and Sastry, Girish and Askell, Amanda and Mishkin, Pamela and Clark, Jack and others},
  booktitle={International conference on machine learning},
  pages={8748--8763},
  year={2021},
  organization={PmLR}
}

@article{mildenhall2021nerf,
  title={Nerf: Representing scenes as neural radiance fields for view synthesis},
  author={Mildenhall, Ben and Srinivasan, Pratul P and Tancik, Matthew and Barron, Jonathan T and Ramamoorthi, Ravi and Ng, Ren},
  journal={Communications of the ACM},
  volume={65},
  number={1},
  pages={99--106},
  year={2021},
  publisher={ACM New York, NY, USA}
}

@article{ho2020denoising,
  title={Denoising diffusion probabilistic models},
  author={Ho, Jonathan and Jain, Ajay and Abbeel, Pieter},
  journal={Advances in neural information processing systems},
  volume={33},
  pages={6840--6851},
  year={2020}
}

@inproceedings{songdenoising,
  title={Denoising Diffusion Implicit Models},
  author={Song, Jiaming and Meng, Chenlin and Ermon, Stefano},
  booktitle={ICLR},
  year={2021}
}

@inproceedings{liu2024syncdreamer,
  title={SyncDreamer: Generating Multiview-consistent Images from a Single-view Image},
  author={Liu, Yuan and Lin, Cheng and Zeng, Zijiao and Long, Xiaoxiao and Liu, Lingjie and Komura, Taku and Wang, Wenping},
  booktitle={ICLR},
  year={2024}
}

@article{wang2023prolificdreamer,
  title={Prolificdreamer: High-fidelity and diverse text-to-3d generation with variational score distillation},
  author={Wang, Zhengyi and Lu, Cheng and Wang, Yikai and Bao, Fan and Li, Chongxuan and Su, Hang and Zhu, Jun},
  journal={Advances in neural information processing systems},
  volume={36},
  pages={8406--8441},
  year={2023}
}

@inproceedings{yu2023text,
  title={Text-to-3D with Classifier Score Distillation},
  author={Yu, Xin and Guo, Yuan-Chen and Li, Yangguang and Liang, Ding and Zhang, Song-Hai and Qi, Xiaojuan},
  booktitle={ICLR},
  year={2023}
}

@inproceedings{long2024wonder3d,
  title={Wonder3d: Single image to 3d using cross-domain diffusion},
  author={Long, Xiaoxiao and Guo, Yuan-Chen and Lin, Cheng and Liu, Yuan and Dou, Zhiyang and Liu, Lingjie and Ma, Yuexin and Zhang, Song-Hai and Habermann, Marc and Theobalt, Christian and others},
  booktitle={Proceedings of the IEEE/CVF conference on computer vision and pattern recognition},
  pages={9970--9980},
  year={2024}
}

@inproceedings{shimvdream,
  title={MVDream: Multi-view Diffusion for 3D Generation},
  author={Shi, Yichun and Wang, Peng and Ye, Jianglong and Mai, Long and Li, Kejie and Yang, Xiao},
  booktitle={ICLR},
  year={2023}
}

@inproceedings{liu2023zero,
  title={Zero-1-to-3: Zero-shot one image to 3d object},
  author={Liu, Ruoshi and Wu, Rundi and Van Hoorick, Basile and Tokmakov, Pavel and Zakharov, Sergey and Vondrick, Carl},
  booktitle={Proceedings of the IEEE/CVF international conference on computer vision},
  pages={9298--9309},
  year={2023}
}

@article{gao2024cat3d,
  title={CAT3D: Create Anything in 3D with Multi-View Diffusion Models},
  author={Gao, Ruiqi and Holynski, Aleksander and Henzler, Philipp and Brussee, Arthur and Martin Brualla, Ricardo and Srinivasan, Pratul and Barron, Jonathan and Poole, Ben},
  journal={Advances in Neural Information Processing Systems},
  volume={37},
  pages={75468--75494},
  year={2024}
}

@article{vaswani2017attention,
  title={Attention is all you need},
  author={Vaswani, Ashish and Shazeer, Noam and Parmar, Niki and Uszkoreit, Jakob and Jones, Llion and Gomez, Aidan N and Kaiser, {\L}ukasz and Polosukhin, Illia},
  journal={Advances in neural information processing systems},
  volume={30},
  year={2017}
}

@inproceedings{li2024instant3d,
  title={Instant3D: Fast Text-to-3D with Sparse-view Generation and Large Reconstruction Model},
  author={Li, Jiahao and Tan, Hao and Zhang, Kai and Xu, Zexiang and Luan, Fujun and Xu, Yinghao and Hong, Yicong and Sunkavalli, Kalyan and Shakhnarovich, Greg and Bi, Sai},
  booktitle={ICLR},
  year={2024}
}

@inproceedings{xu2024dmv3d,
  title={DMV3D: Denoising Multi-view Diffusion Using 3D Large Reconstruction Model},
  author={Xu, Yinghao and Tan, Hao and Luan, Fujun and Bi, Sai and Wang, Peng and Li, Jiahao and Shi, Zifan and Sunkavalli, Kalyan and Wetzstein, Gordon and Xu, Zexiang and others},
  booktitle={ICLR},
  year={2024}
}

@article{huang2025cupid,
  title={CUPID: Generative 3D Reconstruction via Joint Object and Pose Modeling},
  author={Huang, Binbin and Duan, Haobin and Zhao, Yiqun and Zhao, Zibo and Ma, Yi and Gao, Shenghua},
  journal={arXiv preprint arXiv:2510.20776},
  year={2025}
}

@inproceedings{daisvg,
  title={SVG: 3D Stereoscopic Video Generation via Denoising Frame Matrix},
  author={Dai, Peng and Tan, Feitong and Xu, Qiangeng and Futschik, David and Du, Ruofei and Fanello, Sean and QI, XIAOJUAN and Zhang, Yinda},
  booktitle={ICLR},
  year={2025}
}

@article{li2025ss4d,
  title={SS4D: Native 4D Generative Model via Structured Spacetime Latents},
  author={Li, Zhibing and Zhang, Mengchen and Wu, Tong and Tan, Jing and Wang, Jiaqi and Lin, Dahua},
  journal={ACM Transactions on Graphics (TOG)},
  volume={44},
  number={6},
  pages={1--12},
  year={2025},
  publisher={ACM New York, NY, USA}
}

@inproceedings{chen2025artilatent,
  title={ArtiLatent: Realistic Articulated 3D Object Generation via Structured Latents},
  author={Chen, Honghua and Lan, Yushi and Chen, Yongwei and Pan, Xingang},
  booktitle={Proceedings of the SIGGRAPH Asia 2025 Conference Papers},
  pages={1--11},
  year={2025}
}

@article{li2025particulate,
  title={Particulate: Feed-Forward 3D Object Articulation},
  author={Li, Ruining and Yao, Yuxin and Zheng, Chuanxia and Rupprecht, Christian and Lasenby, Joan and Wu, Shangzhe and Vedaldi, Andrea},
  journal={arXiv preprint arXiv:2512.11798},
  year={2025}
}

@article{liu2022flow,
  title={Flow straight and fast: Learning to generate and transfer data with rectified flow},
  author={Liu, Xingchao and Gong, Chengyue and Liu, Qiang},
  journal={arXiv preprint arXiv:2209.03003},
  year={2022}
}

@inproceedings{kingma2013auto,
  title={Auto-encoding variational bayes},
  author={Kingma, Diederik P and Welling, Max},
  booktitle={ICLR},
  year={2014}
}

@article{yang2025latent,
  title={Latent denoising makes good visual tokenizers},
  author={Yang, Jiawei and Li, Tianhong and Fan, Lijie and Tian, Yonglong and Wang, Yue},
  journal={arXiv preprint arXiv:2507.15856},
  year={2025}
}

@article{yao2025towards,
  title={Towards Scalable Pre-training of Visual Tokenizers for Generation},
  author={Yao, Jingfeng and Song, Yuda and Zhou, Yucong and Wang, Xinggang},
  journal={arXiv preprint arXiv:2512.13687},
  year={2025}
}

@article{oquab2024dinov2,
  title={DINOv2: Learning Robust Visual Features without Supervision},
  author={Oquab, Maxime and Darcet, Timoth{\'e}e and Moutakanni, Th{\'e}o and Vo, Huy and Szafraniec, Marc and Khalidov, Vasil and Fernandez, Pierre and Haziza, Daniel and Massa, Francisco and El-Nouby, Alaaeldin and others},
  journal={Transactions on Machine Learning Research Journal},
  year={2024}
}

@InProceedings{Song_2025_CVPR,
    author    = {Song, Chaoyue and Zhang, Jianfeng and Li, Xiu and Yang, Fan and Chen, Yiwen and Xu, Zhongcong and Liew, Jun Hao and Guo, Xiaoyang and Liu, Fayao and Feng, Jiashi and Lin, Guosheng},
    title     = {MagicArticulate: Make Your 3D Models Articulation-Ready},
    booktitle = {Proceedings of the Computer Vision and Pattern Recognition Conference (CVPR)},
    month     = {June},
    year      = {2025},
    pages     = {15998-16007}
}

@inproceedings{deitke2023objaverse,
  title={Objaverse: A universe of annotated 3d objects},
  author={Deitke, Matt and Schwenk, Dustin and Salvador, Jordi and Weihs, Luca and Michel, Oscar and VanderBilt, Eli and Schmidt, Ludwig and Ehsani, Kiana and Kembhavi, Aniruddha and Farhadi, Ali},
  booktitle={Proceedings of the IEEE/CVF conference on computer vision and pattern recognition},
  pages={13142--13153},
  year={2023}
}

@article{deitke2023objaversexl,
  title={Objaverse-xl: A universe of 10m+ 3d objects},
  author={Deitke, Matt and Liu, Ruoshi and Wallingford, Matthew and Ngo, Huong and Michel, Oscar and Kusupati, Aditya and Fan, Alan and Laforte, Christian and Voleti, Vikram and Gadre, Samir Yitzhak and others},
  journal={Advances in Neural Information Processing Systems},
  volume={36},
  pages={35799--35813},
  year={2023}
}

@article{memoli2011gromov,
  title={Gromov--Wasserstein distances and the metric approach to object matching},
  author={M{\'e}moli, Facundo},
  journal={Foundations of computational mathematics},
  volume={11},
  number={4},
  pages={417--487},
  year={2011},
  publisher={Springer}
}

@article{givens1984class,
  title={A class of Wasserstein metrics for probability distributions.},
  author={Givens, Clark R and Shortt, Rae Michael},
  journal={Michigan Mathematical Journal},
  volume={31},
  number={2},
  pages={231--240},
  year={1984},
  publisher={University of Michigan, Department of Mathematics}
}

@inproceedings{liu2021swin,
  title={Swin transformer: Hierarchical vision transformer using shifted windows},
  author={Liu, Ze and Lin, Yutong and Cao, Yue and Hu, Han and Wei, Yixuan and Zhang, Zheng and Lin, Stephen and Guo, Baining},
  booktitle={Proceedings of the IEEE/CVF international conference on computer vision},
  pages={10012--10022},
  year={2021}
}

@inproceedings{peebles2023scalable,
  title={Scalable diffusion models with transformers},
  author={Peebles, William and Xie, Saining},
  booktitle={Proceedings of the IEEE/CVF international conference on computer vision},
  pages={4195--4205},
  year={2023}
}

@article{Xu2019PredictingAS,
title={Predicting Animation Skeletons for 3D Articulated Models via Volumetric Nets},
author={Zhan Xu and Yang Zhou and Evangelos Kalogerakis and Karan Singh},
journal={2019 International Conference on 3D Vision (3DV)},
year={2019},
pages={298-307},
url={https://api.semanticscholar.org/CorpusID:201309034}
}

@article{RigMesh,
author = {Boros\'{a}n, P\'{e}ter and Jin, Ming and DeCarlo, Doug and Gingold, Yotam and Nealen, Andrew},
title = {RigMesh: automatic rigging for part-based shape modeling and deformation},
year = {2012},
issue_date = {November 2012},
publisher = {Association for Computing Machinery},
address = {New York, NY, USA},
volume = {31},
number = {6},
issn = {0730-0301},
url = {https://doi.org/10.1145/2366145.2366217},
doi = {10.1145/2366145.2366217},
journal = {ACM Trans. Graph.},
month = nov,
articleno = {198},
numpages = {9}
}

@article{Baerentzen14, Title = {Interactive Shape Modeling using a Skeleton-Mesh Co-Representation}, Author = {J. A. B{\ae}rentzen and R. Abdrashitov and K. Singh}, Journal = {ACM Transactions on Graphics (proceedings of ACM SIGGRAPH)}, Number = {4}, Volume = {33}, Year = {2014}}

@inproceedings{FEQE,
author = {Pandey, Karran and B\ae{}rentzen, J. Andreas and Singh, Karan},
title = {Face Extrusion Quad Meshes},
year = {2022},
isbn = {9781450393379},
publisher = {Association for Computing Machinery},
address = {New York, NY, USA},
url = {https://doi.org/10.1145/3528233.3530754},
doi = {10.1145/3528233.3530754},
booktitle = {ACM SIGGRAPH 2022 Conference Proceedings},
articleno = {10},
numpages = {9},
location = {Vancouver, BC, Canada},
series = {SIGGRAPH '22}
}
}

\end{document}